\documentclass[12pt]{article}

\usepackage{amsmath}
\usepackage{amsfonts}
\usepackage{graphicx}
\usepackage{caption}
\usepackage{subcaption}

\usepackage{authblk}
\usepackage{amssymb}

\usepackage{epstopdf}
\usepackage{epsfig}
\usepackage{hyperref}

\usepackage{color}

\date{}

\begin{document}

\title{Quasinormal modes of compact objects in alternative theories of gravity}

\author[1]{Jose Luis Bl\'azquez-Salcedo \thanks{\href{mailto:jose.blazquez.salcedo@uni-oldenburg.de}{jose.blazquez.salcedo@uni-oldenburg.de}}}

\author[1]{Zahra Altaha Motahar \thanks{\href{mailto:zahra.motahar@uni-oldenburg.de}{zahra.motahar@uni-oldenburg.de}}}

\author[2,3]{Daniela D. Doneva
\thanks{\href{mailto:daniela.doneva@uni-tuebingen.de }{daniela.doneva@uni-tuebingen.de }}}

\author[4]{Fech Scen Khoo \thanks{\href{mailto:fech.scen.khoo@irb.hr}{fech.scen.khoo@irb.hr}}} 

\author[1]{Jutta Kunz \thanks{\href{mailto:jutta.kunz@uni-oldenburg.de}{jutta.kunz@uni-oldenburg.de}}} 

\author[5]{Sindy Mojica\thanks{\href{mailto:sindroc@gmail.com}{sindroc@gmail.com}}}

\author[6]{Kalin V. Staykov \thanks{\href{mailto:kstaykov@phys.uni-sofia.bg}{kstaykov@phys.uni-sofia.bg}}}

\author[2,6,7]{Stoytcho S. Yazadjiev \thanks{\href{mailto:yazad@phys.uni-sofia.bg}{yazad@phys.uni-sofia.bg}}}

\affil[1]{Institut f\"ur  Physik, Universit\"at Oldenburg, Postfach 2503,
D-26111 Oldenburg, Germany}

\affil[2]{Theoretical Astrophysics, Eberhard Karls University of T\"ubingen, T\"ubingen 72076, Germany}

\affil[3]{INRNE - Bulgarian Academy of Sciences, 1784  Sofia, Bulgaria}

\affil[4]{Division of Theoretical Physics, Rudjer Bo\v skovi\'c Institute
Bijeni\v cka 54, 10000 Zagreb, Croatia}

\affil[5]{Escuela de F\'isica, Universidad Industrial de Santander,
Bucaramanga 680002, Colombia}

\affil[6]{Department of Theoretical Physics, Faculty of Physics, Sofia University, Sofia 1164, Bulgaria}

\affil[7]{Institute of Mathematics and Informatics, Bulgarian Academy of Sciences, Acad. G. Bonchev Street 8, Sofia 1113, Bulgaria}

\maketitle

\begin{abstract}

We address quasinormal modes of compact objects 
in several alternative theories of gravity. 
In particular, we focus on black holes and neutron stars with scalar hair.
We consider black holes in dilaton-Einstein-Gau\ss -Bonnet theory, 
and in a generalized scalar-Einstein-Gau\ss -Bonnet theory.
In the latter case scalarized black holes arise,
and we study the stability of the different branches of solutions. 
In particular, we discuss how the spectrum of quasinormal modes 
is changed by the presence of a non-trivial scalar field outside 
the black hole horizon.
We discuss the existence of an (effective) minimum mass in these models, 
and how the spectrum of modes becomes richer 
as compared to general relativity, when a scalar field is present.
Subsequently we discuss the effect of scalar hair 
for realistic neutron star models. 
Here we consider $R^2$ gravity, scalar-tensor theory, 
a particular subsector of Horndeski theory 
with a non-minimal derivative coupling, 
and again dilatonic-Einstein-Gau\ss -Bonnet theory. 
Because of the current lack of knowledge on the internal composition 
of the neutron stars, we focus on universal relations 
for the quasinormal modes, that are largely independent
of the equations of state and thus the matter content of the stars.

\end{abstract}

\pagebreak

\tableofcontents

\section{Introduction}

The discovery of gravitational waves has led to a new era in physics
and astronomy
\cite{Abbott:2016blz,Abbott:2016nmj,Abbott:2017oio,Abbott:2017vtc,
Abbott:2017gyy,TheLIGOScientific:2017qsa,GBM:2017lvd,Coulter:2017wya},
where multi-messenger observations including and triggered by gravitational waves
promise to elucidate some of the most intriguing and at the same time
cataclysmic events in nature.

The emission of gravitational waves as produced by 
colliding black holes and neutron stars in the observed events
follows a typical sequence of phases, consisting of inspiral, merger and ringdown.
The analysis of the gravitational wave signals received by LIGO/VIRGO
has been based on templates obtained on the basis of general relativity (GR),
and has led to a good first understanding of the objects involved in the
collisions. In particular, the data can be well interpreted
by assuming that in five observed events Kerr black holes with (at least in part)
rather large masses have collided, merged and formed
even more massive black holes. Similarly, the single remaining event
(published so far) is well described by assuming that a pair of neutron stars 
has merged and led to a kilonova, as subsequently confirmed by observations
in the full range of the electromagnetic spectrum
(see, e.g., \cite{GBM:2017lvd,Coulter:2017wya,Abbott:2018wiz} 
and references therein).

In the future further interferometers will start to operate, 
including KAGRA, LIGO in India, and the Einstein telescope.
In particular, with the new and much more powerful instruments a
considerably more sensitive analysis 
of gravitational waves will be possible,
allowing for the discrimination/exclusion of alternative theories of gravity
as well as of matter models in the form of equations of state (EOSs)
of neutron stars.
Here of particular interest will be the detailed study of the
ringdown phase of the final compact object,
which should allow to extract further information 
on the gravity theory and - in the case of neutron stars -
on the matter composition
(see e.g., \cite{Berti:2015itd,Berti:2018cxi,Berti:2018vdi,Barack:2018yly}).

The ringdown of the final compact object is dominated by its quasinormal modes,
which may be considered as the eigenfunctions of some master equation(s),
describing the first order deviations from an appropriate stationary solution,
such as the Kerr black hole (in the case of GR) or some rotating neutron star
model with a given EOS.
Perturbations of black holes and neutron stars
have been studied for a long time, starting with the work of 
Regge and Wheeler, Zerilli, and Thorne and Campolattaro
\cite{Regge:1957td,Zerilli:1971wd,Thorne:1967a},
and excellent reviews on quasinormal modes of compact objects are, 
for instance, found in
\cite{Kokkotas:1999bd,Nollert:1999ji,Berti:2009kk,Konoplya:2011qq}.
The study of the quasinormal modes of compact objects
does also allow to assess their mode stability.

Whereas quasinormal modes of black holes and neutron stars have
been widely studied in GR by now, much less is known about
quasinormal modes in alternative theories of gravity.
Indeed, various theoretical and experimental reasons indicate
that GR should be modified in strong gravitational fields.
Therefore black holes and neutron stars represent valuable
astrophysical labs to probe gravity in order to affirm, constrain
or discard alternative theories of gravity
\cite{Capozziello:2011et,Will:2014kxa,Berti:2015itd}.

In various alternative theories of gravity the Schwarzschild and the
Kerr solutions of GR also represent solutions of 
the new generalized field equations.
However, for instance, in the presence of scalar fields
in addition so-called scalarized black holes solutions may arise,
as found, e.g., in \cite{Doneva:2017bvd,Silva:2017uqg,Antoniou:2017acq}.
In other alternative theories the Schwarzschild and Kerr solutions may 
no longer be solutions of the new generalized field equations
and only be approached in some limit, while the black hole
solutions of the theory may possess rather different properties,
and, for instance, carry scalar hair
as shown, e.g., in \cite{Mignemi:1992nt,Kanti:1995vq,Kleihaus:2011tg}.

For neutron stars the presence of scalar fields in 
alternative theories of gravity can likewise lead to 
scalarized neutron stars, as demonstrated, e.g., in 
\cite{Damour:1993hw} or \cite{Doneva:2017bvd}.
Alternatively, neutron stars may also carry scalar hair,
as seen, e.g., in \cite{Pani:2011xm,Kleihaus:2016dui},
while the GR neutron star solutions are no longer
solutions of the generalized field equations.
However, for neutron stars their unknown composition
and thus EOS represents a further complication
when trying to exploit their astrophysical signatures
to learn about alternative theories of gravity.

Here a rather useful tool may be provided by the various
universal relations of neutron stars
(see e.g., the recent reviews \cite{Yagi:2016bkt,Doneva:2017jop}).
Whereas neutron star models, and thus neutron star structure 
and quasinormal modes,
depend typically strongly on the specific choice of EOS,
there are various universal relations, i.e.,
almost EOS-independent relations,
between properly scaled neutron star properties.
Such universal relations arise not only for neutron stars in GR,
but also for neutron stars in various alternative theories of
gravity \cite{Yagi:2016bkt,Doneva:2017jop}.

While there is a large variety of alternative theories of gravity,
we will here focus on a subset of theories,
which all either contain a gravitational scalar field
like scalar-tensor theories
\cite{Brans:1961sx,Damour:1992we,Damour:1996ke,Fujii:2003pa},
extended scalar theories
\cite{Doneva:2017bvd,Silva:2017uqg,Antoniou:2017acq},
Horndeski theories
\cite{Horndeski:1974wa,Nicolis:2008in,Kobayashi:2011nu},
and string theory motivated dilatonic theories
\cite{Gross:1986mw,Metsaev:1987zx},
or alternative gravity theories
which can be reformulated in terms of a scalar field
like $f(R)$ theories
\cite{Sotiriou:2008rp,DeFelice:2010aj,Capozziello:2011et}.

In all cases, we will discuss the quasinormal modes of
black holes or neutron stars, considering mostly
axial quasinormal modes for simplicity,
when the much more involved calculation of the 
polar modes has not yet been done or is still in progress.
We will see that the presence of the scalar field 
can lead to significant changes in the mode spectrum.

In section 2 we will provide the general formalism 
for calculating quasinormal modes, and we will
describe a set of appropriate numerical methods.
In section 3 
we will discuss the quasinormal modes of black holes
in scalar-Einstein-Gau\ss -Bonnet (sEGB) theory,
addressing first dilatonic black holes and then
scalarized black holes.
The quasinormal modes of neutron stars will be
presented in section 4,
considering $R^2$ gravity, scalar-tensor theory (STT),
Horndeski gravity with non-minimal derivative coupling,
and dilatonic Einstein-Gau\ss -Bonnet (dEGB) theory.
We will conclude in section 5.

\section{Quasinormal modes: general formalism} \label{sec_QNM}

In this section we address the general formalism to
obtain the quasinormal modes of black holes and neutron stars,
We start by considering the static spherically symmetric
background configurations of these compact objects.
Subsequently we consider their first order perturbations 
and present, in particular, the decompositions of the
perturbations in the axial and polar channels.
We discuss the asymptotic behaviour of the solutions,
and present numerical methods to solve for the modes,
focussing on the ``shooting'' method and exterior complex scaling.

\subsection{The backgrounds} \label{sec_o0}

Let us start by addressing the general formalism
employed for the static spherically symmetric
background configurations, for which
the perturbations of the various fields 
present in the respective systems then have to be performed.

The gravitational field will be described by a metric, 
which at zeroth order will be static and spherically symmetric,
and denoted by $g^{(0)}$. 
We parametrize this metric by the standard Ansatz
\begin{equation}
ds^{2}=g^{(0)}_{\mu\nu}dx^{\mu}dx^{\nu}=-F(r)dt^{2} + K(r)dr^{2} + r^{2}(d\theta^{2} 
+ \sin^{2}\theta \, d\varphi^{2})~,
\label{ds2_0}
\end{equation} 
where $F(r)$ and $K(r)$ are functions of the radial coordinate $r$, 
associated with the respective background metric.
Since we are interested in astrophysically relevant solutions, 
the asymptotic behaviour of the metric as infinity is approached,
is (in all cases considered) given by 
$\{F,K^{-1}\} \to 1-\frac{2M}{r} + O(r^{-2})$, 
with $M$ determining the total mass of the configuration.

When studying neutron stars we have to take into account the matter 
that composes the star. For all practical porposes we can treat 
the complicated matter inside the star as an effective perfect fluid,
which is parametrized by three quantities: 
the pressure $p$, the density $\rho$ and the four-velocity $u$ of the fluid. 
The corresponding stress energy tensor is 
\begin{equation}
T_{\mu \nu} = (\rho + p)u_{\mu}u_{\nu} + p g_{\mu \nu}.
\label{SE-T}
\end{equation}
which in the spherically symmetric case is simply determined by
\begin{equation}
p = p_0(r), \ \ \ 
\rho = \rho_0(r),  \ \ \
u = u^{(0)} = u^t \partial_t 
\label{p_rho_u_0}
\end{equation}
with $u^2=-1$.

To describe the nuclear matter, we will assume 
that there exists a barotropic EOS,
relating the energy density to the baryon density 
and the pressure of the effective perfect fluid. 
In practice this results in a relation of the form $\rho=\rho(p)$. 
The EOS describing the interior of the neutron stars 
is not well understood, and many models have been proposed 
predicting different profiles for the mass, radius and other properties 
of the stars. In particular, we will address how the different EOSs affect 
the quasinormal mode spectrum.

In the models we will consider, black holes and neutron stars 
will possess a non-trivial background scalar field. This scalar field will 
also be spherically symmetric, and be parametrized by a single function,
\begin{equation}
\phi = \phi_0(r).
\label{phi0}
\end{equation} 
We will focus on solutions with zero cosmological value 
of the scalar field, $\phi|_{\infty} = 0$. 
In general the scalar field decays as $\phi_0 \sim D/r$, 
with $D$ the scalar charge of the configuration. However, if the scalar field 
is massive (as it happens in the case of $R^2$ gravity), 
the field decays exponentially.

In practice, the previous Ansatz for the fields 
that compose the configurations
reduces the field equations of the respective action
to a system of ordinary differential equations. 
This is a system for $F$, $K$, $\phi_0$ -- 
and, in the case of neutron stars, also $p_0$. 
These functions have to be determined by integrating the field equations 
subject to certain boundary conditions. 
These conditions are essentially determined by demanding
the presence of a regular event horizon in the case of black holes, 
and a regular center and a surface in the case of neutron stars,
as well as asymptotic flatness in both cases.
The lack of analytical solutions means that we have to integrate 
these equations numerically. 
We will give more details on the particular properties 
of the background solutions of each model in the following sections.

\subsection{Non-radial perturbations} \label{sec_pert}

Once the static and spherically symmetric configuration is known, 
we can employ perturbation theory 
\cite{Thorne:1967a,Thorne:1969a,Thorne:1969b,Thorne:1969rba,Lindblom:1983ps,Detweiler:1985zz,Chandrasekhar:1991fi,Chandrasekhar:1991a,Chandrasekhar:1991b,Ipser:1991ind,Kojima:1992ie}. 
Let $\epsilon<<1$ be the parameter controlling the order of the perturbation. 
Then the metric field can be perturbed by generic non-radial perturbations 
up to second order in $\epsilon$:
\begin{equation}
g_{\mu\nu} = g_{\mu\nu}^{(0)}(r) + \epsilon h_{\mu\nu}(t,r,\theta,\varphi)~,
\label{metric_pert}
\end{equation}
where $h_{\mu\nu}(t,r,\theta,\varphi)$ is the metric perturbation.
In the presence of matter, we should perturb also the different fluid quantities:
\begin{eqnarray}
p = p_0(r) + \epsilon \delta p(t,r,\theta,\varphi)~, \\
\rho = \rho_0(r) + \epsilon \delta \rho(t,r,\theta,\varphi)~, \\
u = u^{(0)} + \epsilon \delta u(t,r,\theta,\varphi)~. 
\label{matter_pert}
\end{eqnarray}
Finally, the scalar field should also be perturbed up to order $\epsilon$:
\begin{eqnarray}
\phi = \phi_{0}(r) + \epsilon \delta \phi(t,r,\theta,\varphi)~.
\label{Phi_pert}
\end{eqnarray}

When plugging this Ansatz for generic perturbations into the field equations, 
and truncating them at linear order in $\epsilon$, 
one obtains a system of partial differential equations 
for the perturbation functions. 
This system, although linearized around the background configuration, 
is very complicated to study. 

In order to simplify the problem further, 
we decompose the angular dependence into spherical tensor harmonics. 
This is a base of tensors and vectors that allow to decompose 
the angular dependence in terms of the spherical harmonic functions 
$Y_{lm}(\theta,\varphi)$. 
This decomposition takes advantage of the spherical symmetry 
of the background configuration, which allows for the complete decoupling 
of the angular dependence of the solution. 
However, the field equations will possess an explicit dependence 
on the angular number $l$.
(They do not depend on the magnetic number $m$ because of the symmetry.) 
The spherical symmetry allows to decouple the perturbation equations 
into two independent channels: 
axial modes, with perturbations possessing odd-parity 
under reflection of the angular coordinates 
($Y_{lm}(\theta,\varphi) \rightarrow Y_{lm}(\pi-\theta,\pi+\varphi) =
(-1)^{l+1}Y_{lm}(\theta,\varphi)$), 
and polar modes, with even-parity 
($Y_{lm}(\theta,\varphi) \rightarrow Y_{lm}(\pi-\theta,\pi+\varphi) =
(-1)^{l}Y_{lm}(\theta,\varphi)$).

Moreover, instead of studying the full time dependent problem, 
we can perform a Laplace transformation of the fields. 
In practice this is equivalent to assume a time dependence 
of the form $e^{-i \omega t}$, i.e., a mode decomposition. 
The resulting field equations do not depend on time any more, 
but the eigenvalue $\omega$ is introduced explicitly into the equations. 

The eigenvalue $\omega$, in general, possesses a real and an imaginary part, 
$\omega=\omega_R + i \omega_I$ . The frequency of the ringdown 
is given by $\omega_R$, while the imaginary part determines if the perturbation 
is stable or unstable. 
Stable perturbations are characterized by $\omega_I<0$, 
meaning the mode decays exponentially with time,
and the decay time is $\tau = -1/\omega_I$. 
Unstable perturbations have $\omega_I>0$, and explode exponentially with time. 
We will see the importance of this distinction for the asymptotic behaviour 
of the perturbations in the next subsection.
Let us now give explicitly the Ansatz describing the different perturbations 
of the fields. 

\begin{itemize}
\item {Axial channel}
\end{itemize}

We will start with the axial perturbations. These are the simplest ones. 
The metric perturbation is given, in general, by
\begin{equation}
h_{\mu\nu}^{(\text{axial})} =\int d\omega \, \sum\limits_{l,m}
e^{-i\omega t}  \left[
\begin{array}{c c c c}
	0 & 0 & -h_{0}	\frac{1}{\sin\theta}\frac{\partial}{\partial\varphi}Y_{lm} & h_{0}	\sin\theta\frac{\partial}{\partial\theta}Y_{lm} \\
	0 & 0 & -h_{1}	\frac{1}{\sin\theta}\frac{\partial}{\partial\varphi}Y_{lm}  
	& h_{1}	\sin\theta\frac{\partial}{\partial\theta}Y_{lm} \\
-h_{0}	\frac{1}{\sin\theta}\frac{\partial}{\partial\varphi}Y_{lm} & 
	-h_{1}	\frac{1}{\sin\theta}\frac{\partial}{\partial\varphi}Y_{lm} 
	& h_2	\frac{1}{2\sin\theta}X_{lm}  & -\frac{1}{2}h_2 \sin\theta \, W_{lm}
 \\
h_{0}	\sin\theta\frac{\partial}{\partial\theta}Y_{lm} & h_{1}	\sin\theta\frac{\partial}{\partial\theta}Y_{lm}  
   & -\frac{1}{2}h_2 \sin\theta \, W_{lm} & -\frac{1}{2} h_2	\sin\theta \, X_{lm}	
\end{array}
\right]
\end{equation}
where we have defined
\begin{equation}
X_{lm} = 2\frac{\partial^2}{\partial\theta\,\partial\varphi}Y_{lm} - 2 \cot\theta \frac{\partial}{\partial\varphi}Y_{lm}~, \quad
W_{lm}
= \frac{\partial^2}{\partial\theta^2}Y_{lm} - \cot\theta \frac{\partial}{\partial\theta}Y_{lm}
-\frac{1}{\sin^2\theta}\frac{\partial^2}{\partial\varphi^2}Y_{lm}~.
\end{equation}
The functions $h_{0}, h_{1},h_2$ depend on $r$, the angular numbers $l$, $m$ 
and the complex frequency $\omega$. 
The Ansatz can be simplified by choosing the Regge-Wheeler gauge with $h_2=0$.

In the case of neutron stars it is possible to choose the gauge such that 
the perturbation of the four-velocity is trivial
\begin{equation}
\delta {u}_{\mu}^{(\text{axial})}= 0.
\label{u_axial}
\end{equation}

The axial channel is simpler because it does not couple to scalar perturbations. 
This means that in this channel 
\begin{equation}
\delta \rho= \delta p= \delta \phi=0 \, . 
\end{equation}
Essentially, the axial perturbations give rise to purely gravitational radiation. 
We will see that the axial quasinormal modes are in general more insensitive 
to the matter content and presence of the scalar field than the polar modes.

\begin{itemize}
\item {Polar channel}
\end{itemize}

The polar sector is more complicated. The Ansatz for the metric perturbations 
can be written as
\begin{equation}
h_{\mu\nu}^{(\text{polar})} = \int d\omega \, \sum\limits_{l,m}
e^{-i\omega t}  \left[
\begin{array}{c c c c}
	2N F Y_{lm} & -H_1Y_{lm} & 
	-h_{0p}\frac{\partial}{\partial\theta}Y_{lm} & -h_{0p}\frac{\partial}{\partial\varphi} Y_{lm} \\
	-H_1Y_{lm} & -2K L Y_{lm} & 
	h_{1p}\frac{\partial}{\partial\theta}Y_{lm} & h_{1p}\frac{\partial}{\partial\varphi}Y_{lm} \\
	-h_{0p}\frac{\partial}{\partial\theta}Y_{lm} & h_{1p} \frac{\partial}{\partial\theta}Y_{lm} 
	& B & -r^2VX_{lm}
	\\
	 -h_{0p} \frac{\partial}{\partial\varphi} Y_{lm} & h_{1p}\frac{\partial}{\partial\varphi} Y_{lm}  & 
  -r^2 V X_{lm} & A \\
\end{array}
\right]~,
\end{equation}
where we have defined
$A = (l(l+1)V - 2 T)r^2 \sin^2\theta \, Y_{lm} + r^2 V \sin^2\theta\,W_{lm}$ and 
$B = (l(l+1)V - 2 T)r^2  Y_{lm} - r^2 V W_{lm}$. 
Again, the perturbation functions $N, V, T, L, H_1, h_{0p}, h_{1p}$ 
depend on the radial coordinate $r$, the angular numbers $l$, $m$, 
and the complex frequency $\omega$. 
Similarly, we can choose a convenient gauge with $h_{0p}=h_{1p}=V=0$. 

In the case of neutron stars, a suitable Ansatz for the perturbation 
of the four-velocity is given by 
\begin{equation}
\delta {u}^{\mu}_{(\text{polar})}= \int d\omega \, \sum\limits_{l,m} e^{-i\omega t}  \left(\frac{-N}{\sqrt{-F}}Y_{lm},W_f Y_{lm},V_f \frac{\partial}{\partial\theta} Y_{lm},V_f \frac{\partial}{\partial\varphi} Y_{lm}\right) \ ,
\label{u_polar}
\end{equation}
where we have introduced the perturbation functions $W_f$ and $V_f$. 
But the polar channel includes scalar perturbations, which means that
we have to perturb the energy density and the pressure of the fluid 
composing the neutron star, as well,
\begin{eqnarray} 
\delta \rho^{(\text{polar})} =  \int d\omega \, \sum_{l,m} e^{-i\omega t} \rho_1 \, Y_{lm}~, \\
\delta p^{(\text{polar})} =  \int d\omega \,\sum_{l,m}  e^{-i\omega t} p_1 \, Y_{lm}~.
\end{eqnarray}
Finally, the scalar field possesses a similar decomposition,
\begin{equation} 
\delta \phi^{(\text{polar})} =  \int d\omega \, \sum_{l,m} e^{-i\omega t} \Phi_1 \, Y_{lm}~,
\end{equation}
where we have introduced the function $\Phi_1$ that depends on $r$, $l$, $m$ 
and $\omega$, like the metric perturbation functions.

With the previous Ansatz for axial and polar perturbations 
we can simplify the field equations to order $\epsilon$. 
In practice, after some lengthy algebra, 
the equations describing the perturbations can be written as
\begin{equation}
\frac{d}{dr}{{\Psi}}_{(i)}+{U}_{(i)}{\Psi}_{(i)}=0~,
\label{eq_perturbations}
\end{equation}
where $(i)=axial$, $polar$. 
The matrix $U_{(i)}$ contains the coefficients of the equations, 
which are given by combinations of the functions 
of the static background metric ($F(r), K(r)$), 
matter distribution ($p_0(r), \rho_0(r)$) and scalar field ($\phi_0(r)$), 
the angular number $l$ 
(there is degeneracy with respect to $m$), 
and the frequency eigenvalue $\omega$. 

In the axial case $\Psi_{axial} = (h_0, h_1)$. 
This is a system of two coupled first order differential equations. 
The equations can be combined into a single second order differential equation, 
a generalized version of the Regge-Wheeler equation of GR.

In the polar case $\Psi_{polar} = (H_1, T, \Phi_1, \frac{d}{dr}\Phi_1)$ 
for black holes or in the exterior region of neutron stars.
This is a system of four coupled first order differential equations, 
which can be rewritten in terms of two coupled second order equations. 
In the GR limit these two equations decouple, 
resulting in the Zerilli equation for the gravitational perturbations, 
and the test field wave equation for the scalar perturbations. 

Inside a neutron star
$\Psi_{polar} = (H_1, T, p_1, V_f, \Phi_1, \frac{d}{dr}\Phi_1)$, which
results in a complicated coupled system of six ordinary differential equations
that includes the pressure fluctuations.  
Thus the polar modes are expected to present 
a richer quasinormal mode spectrum than the axial modes 
in the presence of matter and/or non-trivial scalar fields, 
since they couple to the fluctuations of these extra fields.

\subsection{Asymptotic behaviour} \label{sec_asymp}

Generically, an asymptotic study of the Eq.~(\ref{eq_perturbations}) 
for $r\to\infty$ reveals that a perturbation can be expressed
as a superposition of two wave solutions,
\begin{eqnarray} %
\Psi \sim A_{in} e^{-i\omega (t+R^*)} + A_{out} e^{-i\omega (t-R^*)},
\end{eqnarray}
where $A_{in}$ is the amplitude of the wave falling into the object, 
and $A_{out}$ is the amplitude of the wave being radiated away from the object. 
The coordinate $R^*$ is defined as a generalized tortoise coordinate, 
which asymptotically coincides with the standard GR tortoise coordinate.

Since we are interested in studying the resonant frequencies 
that characterize the ringdown phase of gravitational waves 
radiated away from black holes and neutron stars, 
we should consider only solutions that asymptotically have $A_{in}=0$.
In practice this means that the radial part of the perturbation functions 
introduced possesses the following behaviour:
\begin{eqnarray}
\Psi \xrightarrow[r \to \infty]{} e^{i\omega R^*} \sim \sin{(\omega_R R^*)} e^{-\omega_I R^*}.
\end{eqnarray}

Obviously stable and unstable perturbations display a very different 
asymptotic behaviour. Unstable perturbations possess $\omega_I>0$, 
which means that the perturbation functions $\Psi$ decay to zero 
as $r$ goes to infinity. 
However, stable perturbations with $\omega_I<0$ possess 
an exponentially divergent behaviour in the perturbation functions. 
Not only that, but any mode with a non-trivial real part 
will show an infinite number of oscillations as $r$ goes to infinity. 
Both the divergent and the oscillatory behaviour 
are challenging in terms of the numerical study of these modes, 
which we will discuss in the next subsection.

In the case of black holes, the nature of the perturbations 
has to be imposed also near the horizon. 
Gravitational waves should be infalling into the horizon,
which means that emission from the black hole 
or a white hole scenario are excluded. 
In the case of neutron stars there is no horizon. 
However, the perturbation functions have to be regular 
at the center of the star, and continuous across the star's surface. 

In any case, it is always possible to perform a perturbative analysis 
of the perturbation functions close to the boundaries. 
In order to do so, one needs a perturbative description 
of the background solution close to the horizon/center and close to infinity. 
With this perturbative solution, it is possible 
to obtain a perturbative description of the matrix $U_{(i)}$ 
of Eq.~(\ref{eq_perturbations}) close to the boundaries. 
Then one can obtain a perturbative approximation 
of the functions $\Psi_{(i)}$. 
This analytical approximation of the perturbation is very useful 
for the numerical analysis of the modes, as we will see in the next subsection. 

\subsection{Numerical methods} \label{sec_num}

Let us now briefly describe the numerical methods we use 
in order to calculate the quasinormal modes. 
The first step is always the calculation of the background solution. 
The solutions studied here have to be generated numerically, 
since no analytical solution is known for the models under consideration. 

To integrate the equations
we make use of Colsys \cite{Ascher:1979iha},
a package that allows to numerically solve boundary value problems 
for systems of ordinary differential equations. 
It implements a spline collocation method, 
which automatically calculates and adapts the mesh points 
in order to achieve a certain required precision on the functions. 
We integrate the equations employing a compactified coordinate, $x$. 
For example, for black holes we take $x=1-r_H/r$, 
while for neutron stars we choose $x=r/(r+r_S)$, 
with $r_H$ and $r_S$ the location of the black hole horizon 
and surface of the star, respectively. 
The compactification is useful because it allows 
to impose boundary conditions exactly at infinity. 
Usually the relative precision of the solutions is estimated 
to be better than $10^{-10}$ for a mesh with several thousands of points 
in the compactified coordinate.

Once the background is obtained, we fix the value of $l$ 
and calculate the quasinormal modes, i.e., 
obtain the resonant values of $\omega$,
that satisfy the various regularity and asymptotic conditions 
for the perturbations. 
The results to be discussed in the following sections 
make use of two different methods, which are described 
in the following.

\begin{figure}[t]
     \centering
\includegraphics[width=0.6\textwidth,angle=0]{{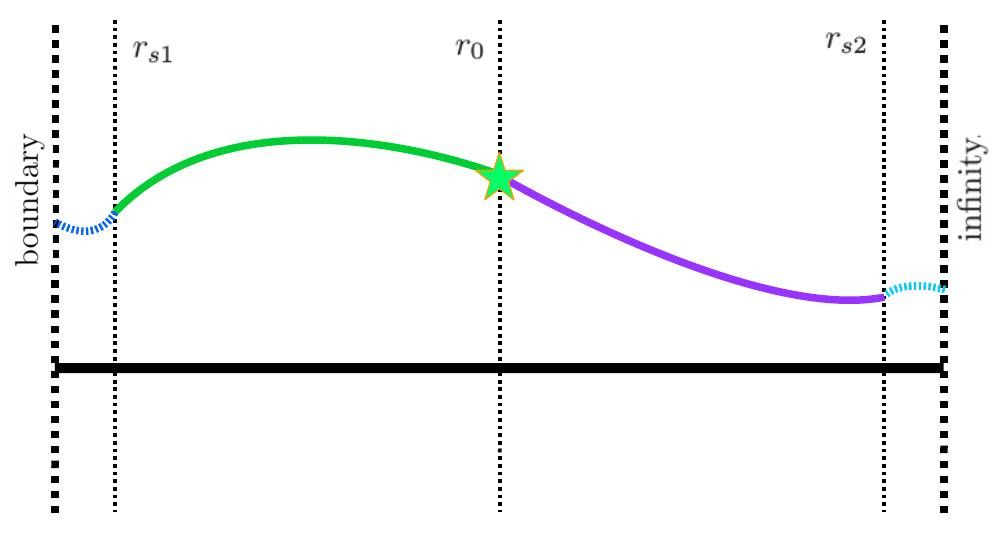}}
         \caption{Schematic example of how the shooting method works. 
The left boundary could be the horizon of a black hole 
or the center of the star. The background is divided into different sections. 
Between the boundary and $r_{s1}$, and infinity and $r_{s2}$, 
we use the perturbative solutions of the perturbation functions 
(short-dashed curves). 
With their help we calculate the initial values 
for the numerical solution of the perturbation functions. 
One solution is then calculated between $r_{s1}$ and $r_0$ (green solid curve), 
and a second solution between $r_0$ and $r_{s2}$ (purple solid curve). 
These two solutions satisfy the required regularity conditions 
and asymptotic behaviour. If $\omega$ is a quasinormal mode, 
then both solutions are continuous at $r_0$.} 
         \label{fig:shooting}
\end{figure}

\begin{itemize}
\item ``Shooting'' method
\end{itemize}

The first method is based on ``shooting'', 
and is schematically represented in Fig.~\ref{fig:shooting}. 
This method makes use of the perturbative solution 
of the perturbation functions that can be obtained close to the boundaries 
i.e., the black hole horizon/center of the star and infinity. 
The perturbative solution is used to describe the perturbation 
in a region extending from the left boundary to a certain value of $r$, 
$r_{s1}$ (short-dashed curve on the left of Fig.~\ref{fig:shooting}). 
We impose on this solution the regularity conditions of the boundary.
We also use the perturbative solution in a region extending from $r_{s2}$ 
up to infinity (short-dashed curve on the right of Fig.~\ref{fig:shooting}).
This solution is required to satisfy the outgoing wave behaviour.

This allows us to calculate initial values for the perturbation equations,
Eq.~(\ref{eq_perturbations}), at $r_{s1}$ and $r_{s2}$. 
We then integrate numerically the equations in a region 
extending from $r_{s1}$ to $r_{0}$, which leads to a first solution 
with regularity at the left boundary 
(green solid curve in Fig.~\ref{fig:shooting}). 
For black holes
typically $r_{s1}/r_H \sim 10^{-5}$. 
Next we integrate numerically the equations in a region 
extending from $r_{0}$ to $r_{s2}$, which leads to a second solution 
with the outgoing wave behaviour at infinity 
(purple solid curve in Fig.~\ref{fig:shooting}). 
For black holes
usually $r_{s2}/r_H \sim 10^{2}$ and $r_0$ is about $(4-6) r_H$. 
However, these parameters can be adjusted in order to test 
and improve the numerical results.

These two numerical solutions should be continuous at $r_0$, 
if $\omega$ is a resonance of the system. 
Indeed, it is found that the solutions are continuous 
(within a required numerical precision) only for discrete 
values of the eigenvalue $\omega$. 
This then determines the spectrum of quasinormal modes. 
To find such values in a systematic way,
we implement a numerical procedure to explore the complex plane 
for $\omega$.
Typically we require $\omega$ to have a relative precision below $10^{-3}$ 
or better.

Because of different sources of error (numerical background, 
limited order of the perturbative solution at the boundaries, 
numerical integration, etc.), it is easy to generate solutions 
with some ingoing wave contamination 
in the right domain of the shooting method. 
This means that in certain cases, we might find 
that the resonant $\omega$ cannot be determined with sufficient precision. 
This potential problem can be avoided in certain cases 
by using the following method.

\begin{itemize}
\item Exterior complex scaling
\end{itemize}

To control the contamination of the numerical solution 
with an ingoing wave component when calculating the resonant frequencies, 
we make use of a method based on exterior complex scaling (see e.g.,
\cite{PhysRevA.44.3060,BlazquezSalcedo:2012pd,Blazquez-Salcedo:2013jka}). 
This method has been successfully employed for axial perturbations, 
where the system reduces to a single second order differential equation 
of the form
\begin{eqnarray}
\frac{d^2 \Psi}{dr^2}+C_0(r)\frac{d\Psi}{dr} + (C_1(r)\omega^2 + C_2(r))\Psi .
\label{eq_Z}
\end{eqnarray}
Instead of integrating Eq.~(\ref{eq_Z}) for the perturbation $\Psi$, 
we can rewrite it in terms of the phase function of the perturbation, 
defining $\frac{dX}{dr} = g X$. 
This gives rise to a Riccati equation \cite{Chandrasekhar:1975zza}
of the form
\begin{eqnarray} 
\label{EQ:phase}
\frac{dg}{dr}+g^2 + C_0(r)g + C_1(r)\omega^2+C_2(r) .
\end{eqnarray}
At infinity, the general solution of Eq.~(\ref{EQ:phase}) behaves as
\begin{eqnarray} \label{phase_inf}
g \sim i\omega \frac{-A_{in} e^{-i\omega r} + A_{out} e^{i\omega r}}{A_{in} e^{-i\omega r} + A_{out} e^{i\omega r}}.
\end{eqnarray}
Asymptotically at infinity the phase function tends to 
$g(r=\infty) = i\omega$, and this happens for purely outgoing waves, 
but also for mixed (ingoing plus outgoing) waves.

The exterior complex scaling method deals with this issue in the following way. 
First an analytical continuation of the equation is made into the complex plane, 
promoting the radial coordinate $r$ to a complex coordinate 
\cite{Andersson:1997xt} and parametrizing it as
\begin{eqnarray} \label{r2y}
r(y) = r_j + y e^{-i\zeta}.
\end{eqnarray}
The parameters $r_j$ and $\zeta$ are constant, 
and we restrict them to be arbitrary positive real numbers. 
The new variable $y$ exists in the interval $y \in [0,\infty)$. 
With this change, the asymptotic behaviour of the phase function $g$
changes when $y \to \infty$, being
\begin{eqnarray} \label{newphase_inf}
g \sim i\omega \frac{-A_{in} e^{-i\xi y} + A_{out} e^{i\xi y}}{A_{in} e^{-i\xi y} + A_{out} e^{i\xi y}},
\end{eqnarray}
with $\xi = \xi_R + i \xi_I$ 
and $\xi_R = \omega_R\cos\zeta + \omega_I\sin\zeta$, 
$\xi_I = \omega_I\cos\zeta -\omega_R\sin\zeta$. 
If we now choose $\zeta$ such that
$\xi_I = \omega_I\cos\zeta -\omega_R\sin\zeta < 0$, 
the condition $g(y=\infty) = i\omega$ enforces $A_{in}=0$, 
and the solution will describe purely outgoing waves.

With this setting, the resulting numerical procedure is slightly different 
from the shooting method. 
Essentially, instead of solving Eq.~(\ref{eq_perturbations}) 
on the right side with $r>r_0$, we obtain the phase function 
on this region using exterior complex scaling 
with boundary condition $g(y=\infty) = i\omega$.

The quasinormal modes are obtained when the left side of the perturbation 
matches at $r=r_0$ with the phase function obtained on the right-hand side. 
Again, this determines within the numerical accuracy 
the discrete set of values of $\omega$. 
We implement a search algorithm that minimizes the difference 
between both solutions at $r=r_j$ using the gradient descent 
by changing the values of $\omega_R$ and $\omega_I$. 
Once we obtain a quasinormal mode, we test the numerical stability 
by changing the auxiliary parameters of the scheme, $r_0$ and $\zeta$. 
Typically the resonances can be obtained with a relative accuracy of $10^{-3}$ 
or better. 

\section{Black holes in scalar-Einstein-Gau\ss -Bonnet gravity} \label{sec_bh}

In this section we will discuss the properties of quasinormal modes 
of black holes in scalar-Einstein-Gau\ss -Bonnet gravity. 
The action of this theory is given by
\begin{eqnarray}
S=&&\frac{1}{16\pi}\int d^4x \sqrt{-g} 
\Big[R + \lambda^2 f(\phi){\cal R}^2_{GB} - 2\nabla_\mu \phi \nabla^\mu \phi 
 \Big] .\label{sEGB_action}
\end{eqnarray}
Here ${\cal R}^2_{GB}$ is the Gau\ss -Bonnet scalar 
(${\cal R}^2_{GB} = 
R_{\mu\nu\rho\sigma}R^{\mu\nu\rho\sigma} -4R_{\mu\nu}R^{\mu\nu} + R^2$), 
$\lambda$ is the Gau\ss -Bonnet coupling constant, 
and $f(\phi)$ is the scalar coupling function. 
If this function is just a constant, 
then the Gau\ss -Bonnet term is topological
and consequently has a trivial contribution to the field equations. 
Otherwise, if the function depends on the scalar field, 
the Gau\ss -Bonnet term enters the equations of motion 
with an extra term in the Einstein equations
\begin{eqnarray}
G_{\mu\nu} = T^{(\phi)}_{\mu\nu} + T^{(GB)}_{\mu\nu} ,\label{sEGB_EE}
\end{eqnarray}
where $G_{\mu\nu}$ is the Einstein tensor and
\begin{eqnarray}
T^{(\phi)}_{\mu\nu} = 2\nabla_\mu \phi \nabla_\mu \phi - g_{\mu\nu}\nabla_\alpha \phi \nabla^\alpha \phi ,\label{Tphi} \\
T^{(GB)}_{\mu\nu} =  2 R \nabla_{(\mu}\psi_{\nu)}-8\nabla^{\alpha}\psi_{(\mu}R_{\nu)\alpha}+4 G_{\mu\nu}\nabla^{\alpha}\psi_{\alpha}+4g_{\mu\nu}R^{\alpha\beta}\nabla_{\alpha}\psi_{\beta} -4R^{\alpha}_{\mu\beta\nu}\nabla^{\beta}\psi_{\alpha},\label{TGB} 
\end{eqnarray}
and we have defined 
$\psi_{\alpha} = \lambda^2 \frac{df}{d\phi} \nabla_{\alpha} \phi$.
The Gau\ss -Bonnet scalar also enters as a source term in the scalar field equation
\begin{eqnarray}
\nabla^2 \phi = -\frac{\lambda^2}{4}\frac{df}{d\phi} {\cal R}^2_{GB}  .
\label{sEGB_sE}
\end{eqnarray}

Employing the Ansatz for spherically symmetric black hole configurations 
(Eqs.~(\ref{ds2_0}) and (\ref{phi0})), 
the field equations simplify to a system of ordinary differential equations. 
With these equations, 
and by requiring the black hole to possess a regular horizon at $r=r_H$, 
the following condition for the existence of black holes with 
a non-trivial scalar field is obtained
\begin{equation}
r_H^4 > 24 \lambda^4 \left(\frac{df}{d\phi}(\phi_{H})\right)^2, 
\label{eq:BC_sqrt_rh}
\end{equation}
where $\phi_{H}$ is the value of the scalar field at the horizon.
We will see that this condition plays an important role 
for several physical properties of the configurations, 
since it essentially translates into a minimum mass for regular black holes. 
However, this minimum mass has to be determined numerically 
for each particular coupling.
In principle, the coupling function $f(\phi)$ can be chosen freely. 
Some cases of special interest will be described in the next sections.

\subsection{Dilatonic coupling} \label{sec_dEGB}

Probably the most studied case in the literature 
corresponds to an exponential coupling function $f(\phi)$ 
containing the scalar field linearly in the exponent. 
This type of coupling arises in the low energy limit
of heterotic string theory with
the scalar field being called dilaton
\cite{Gross:1986mw,Metsaev:1987zx}. 
In the following we will refer to this theory 
as dilatonic Einstein-Gau\ss -Bonnet (dEGB) theory.
The standard coupling of this type can be written as
\begin{equation}
f(\phi)=e^{2\gamma\phi}, \ \ \ \lambda^2=\frac{\alpha}{4},
\label{couling_dEGB}
\end{equation}
where $\gamma$ is a new coupling constant, the dilaton coupling constant, 
which arises in addition to the Gau\ss -Bonnet coupling constant $\alpha$. 
Note, that in the limit of small $\gamma$
a linear coupling is obtained,
since the relevant dilaton Gau\ss -Bonnet (dGB) term of the action 
then reduces to
$\frac{\alpha}{4}e^{2\gamma\phi}{\cal R}^2_{GB} 
\sim \frac{\alpha\gamma}{2}\phi{\cal R}^2_{GB}$.

There are no known closed form solutions for black holes in this theory.
Black hole solutions have been studied perturbatively 
\cite{Mignemi:1992nt,Mignemi:1993ce} 
and numerically \cite{Kanti:1995vq,Torii:1996yi,Guo:2008hf} in the literature 
(see also \cite{Pani:2009wy,Pani:2011gy,Ayzenberg:2014aka,Maselli:2015tta} 
for slowly rotating black holes and 
\cite{Kleihaus:2011tg,Kleihaus:2014lba,Kleihaus:2015aje,Chen:2018jed} 
for rapidly rotating black holes). 
Let us now recall the basic properties of the
static dEGB black holes.

While these dEGB black holes carry a nontrivial dilaton field 
\cite{Kanti:1995vq,Torii:1996yi,Guo:2008hf}, 
the scalar charge $D$ is not a conserved charge of the equations, 
which means that the black holes possess 
what is called secondary scalar hair. 
For fixed values of the coupling constants $\alpha$ and $\gamma$, 
there is a one parameter family of static black holes,
characterized by varying values of the mass $M$ or the horizon area $A_{H}$.

\begin{figure}[t]
     \centering
\includegraphics[width=0.37\textwidth,angle=-90]{{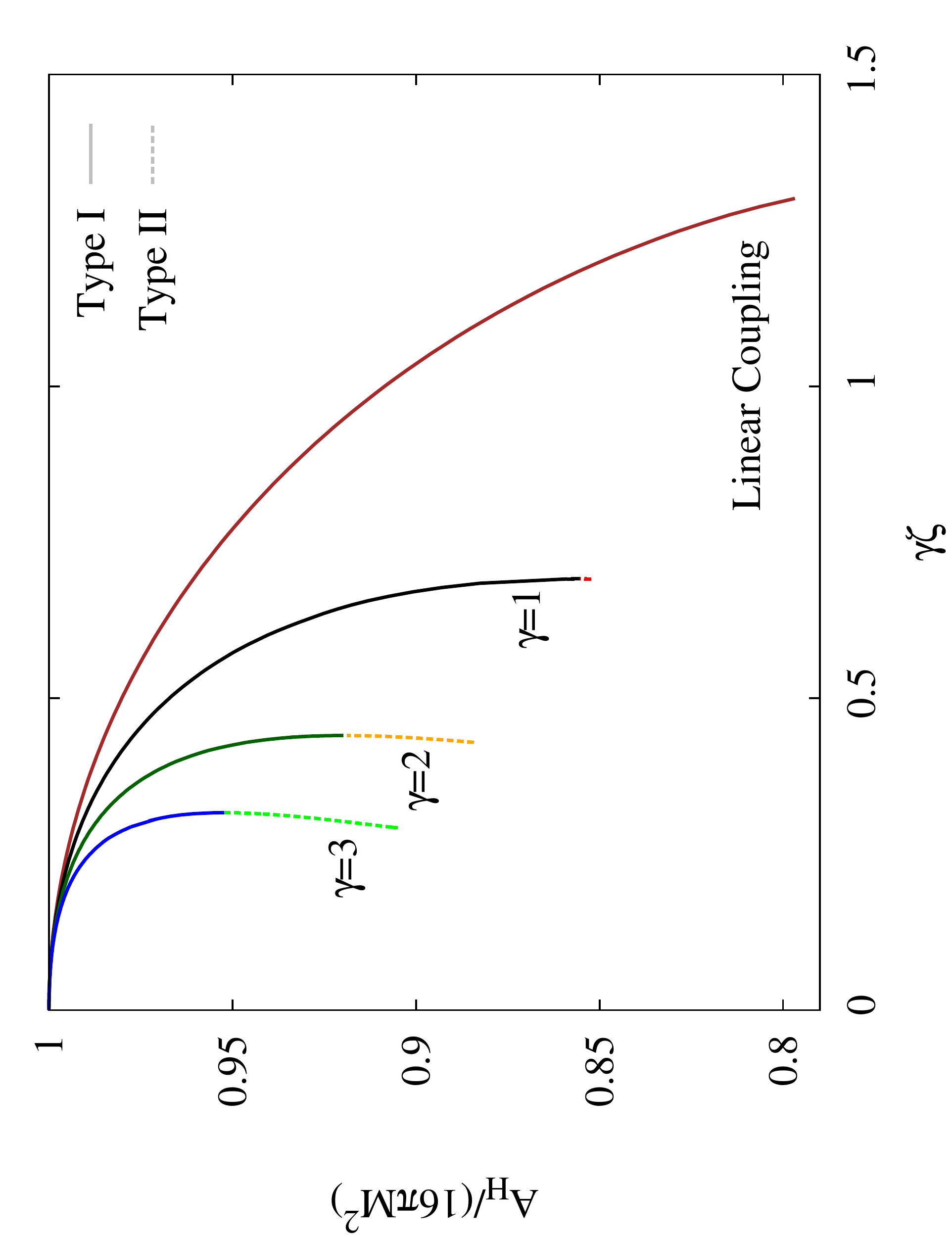}}
\includegraphics[width=0.37\textwidth,angle=-90]{{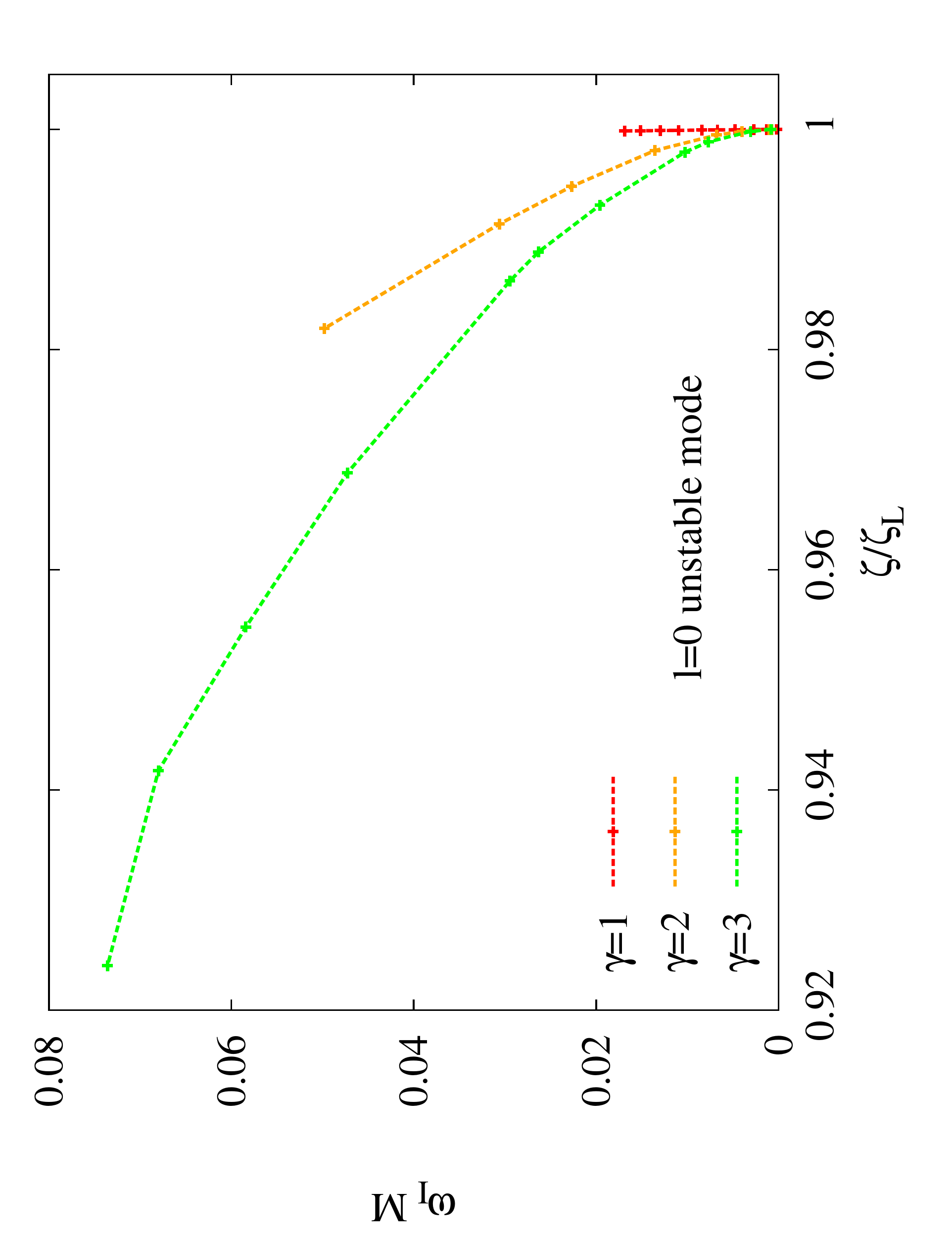}}
         \caption{
(left) Scaled horizon area $A_H$ vs scaled product 
of coupling constants $\gamma\zeta$ for dEGB black holes.
(right) Unstable $l=0$ modes of type II (secondary branch) black holes.}
         \label{fig:dEGB_1}
\end{figure}

The domain of existence of these solutions is bounded 
because of condition (\ref{eq:BC_sqrt_rh}) \cite{Kanti:1995vq}. 
This is seen, for example, in Fig.~\ref{fig:dEGB_1}, where 
the domain of existence of static black holes 
is shown for several values of the
dilaton coupling constant $\gamma$. 
In particular, we show the scaled area of the black hole horizon 
$A_H/16 \pi M^2$ versus the product of the dilaton coupling constant and the
scaled Gau\ss -Bonnet coupling constant, $\gamma\zeta=\gamma\alpha/M^2$.
In the limit $\zeta=0$ the Schwarzschild black hole is reached, 
which satisfies $A_H/16 \pi M^2=1$ and carries no scalar hair.

For a fixed value of $\gamma$ the area decreases 
as the scaled Gau\ss -Bonnet constant $\zeta$ increases. 
The branch of solutions ends at a maximal value of this parameter, $\zeta_L$. 
In particular, it can be seen that increasing $\gamma$ 
decreases the size of the domain of existence, 
since the value of $\zeta_L$ decreases with increasing $\gamma$. 
This limiting value of $\zeta$ is actually related 
to a minimum value of the black hole mass. 
For a fixed value of $\alpha$, the minimum mass allowed for a regular black hole 
is $M_{\rm min}=\sqrt{\alpha/\zeta_L}$. 
Note that for a fixed value of $\alpha$, the larger the mass, 
the more the dEGB black hole resembles a Schwarzschild black hole, 
retaining just very thin scalar hair. 
In contrast, the closer the dEGB black holes is to the minimum mass, 
the more the scalar hair grows.

Interestingly, for larger values of $\gamma$ ($\gamma>0.913$),
a secondary branch of dEGB black holes appears 
close to the limiting value $\zeta_L$
\cite{Kanti:1995vq,Torii:1996yi,Guo:2008hf,Pani:2009wy}. 
We call the black holes on this branch type II black holes. 
In Fig.~\ref{fig:dEGB_1}(left) 
we mark these type II black holes by dashed lines. 
Note, that they are not present in the linear coupling limit. 
When present, the type II black holes extend from $\zeta_L$ 
to some critical value $\zeta_C<\zeta_L$. 
At $\zeta_C$ condition (\ref{eq:BC_sqrt_rh}) is no longer satisfied.

The existence of the type II branches of dEGB black holes means 
that uniqueness of the solutions is lost. 
Thus for fixed values of the coupling constants 
$\alpha$ and $\gamma>0.913$, 
it is possible to find two different black hole solutions 
with the same values of the total mass. 
Indeed, as Fig.~\ref{fig:dEGB_1}(left) indicates,
these black holes possess different horizon properties (different areas),
and they possess different scalar charges.

With these dEGB black hole solutions as backgrounds, 
the formalism and methods described in the previous sections 
can be applied to study their spectrum of quasinormal modes. 
This analysis has been performed in 
\cite{Blazquez-Salcedo:2016enn,Blazquez-Salcedo:2017txk}. 
Let us start by discussing the stability of these solutions.

Type I black holes seem to be free of unstable modes. 
However type II black holes possess an unstable mode for $l=0$. 
This mode starts as a zero mode for the configuration at $\zeta_L$ 
(where the type II black holes bifurcate from the type I black holes). 
The instability then grows as $\zeta$ decreases towards $\zeta_C$. 
We show the eigenvalues of these unstable modes in Fig.~\ref{fig:dEGB_1}(right). 
The unstable modes are present in all type II black holes 
for any value of $\gamma$ considered. 
The results suggest that, 
under full time evolution of a small perturbation of a type II black hole, 
this configuration will develop a radial instability, 
radiating away some amount of scalar hair, 
and presumably reaching a stable type I configuration 
after a time $t \sim 1/\omega_I$.

\begin{figure}[p!]
     \centering
\includegraphics[width=0.38\textwidth,angle=-90]{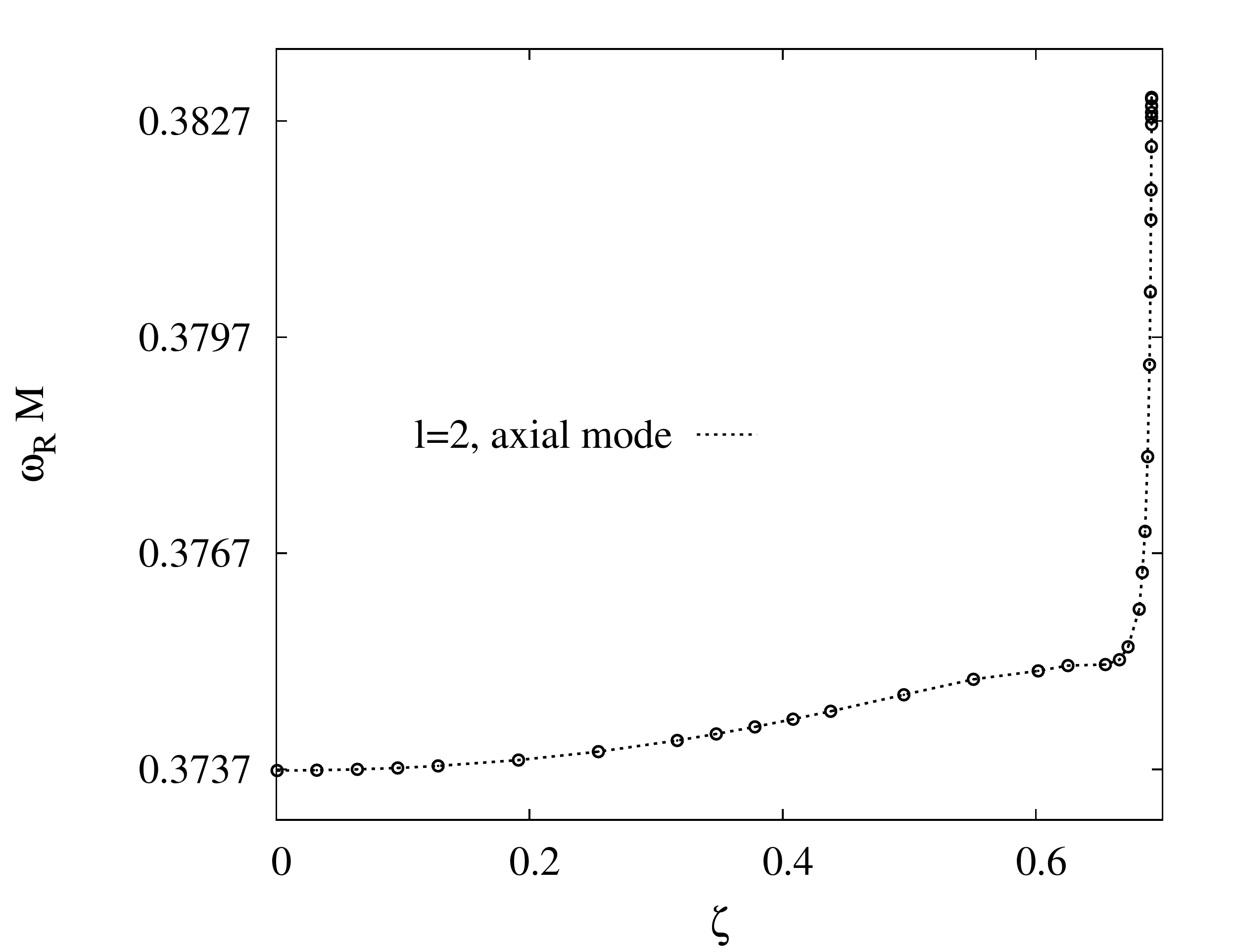}
\includegraphics[width=0.38\textwidth,angle=-90]{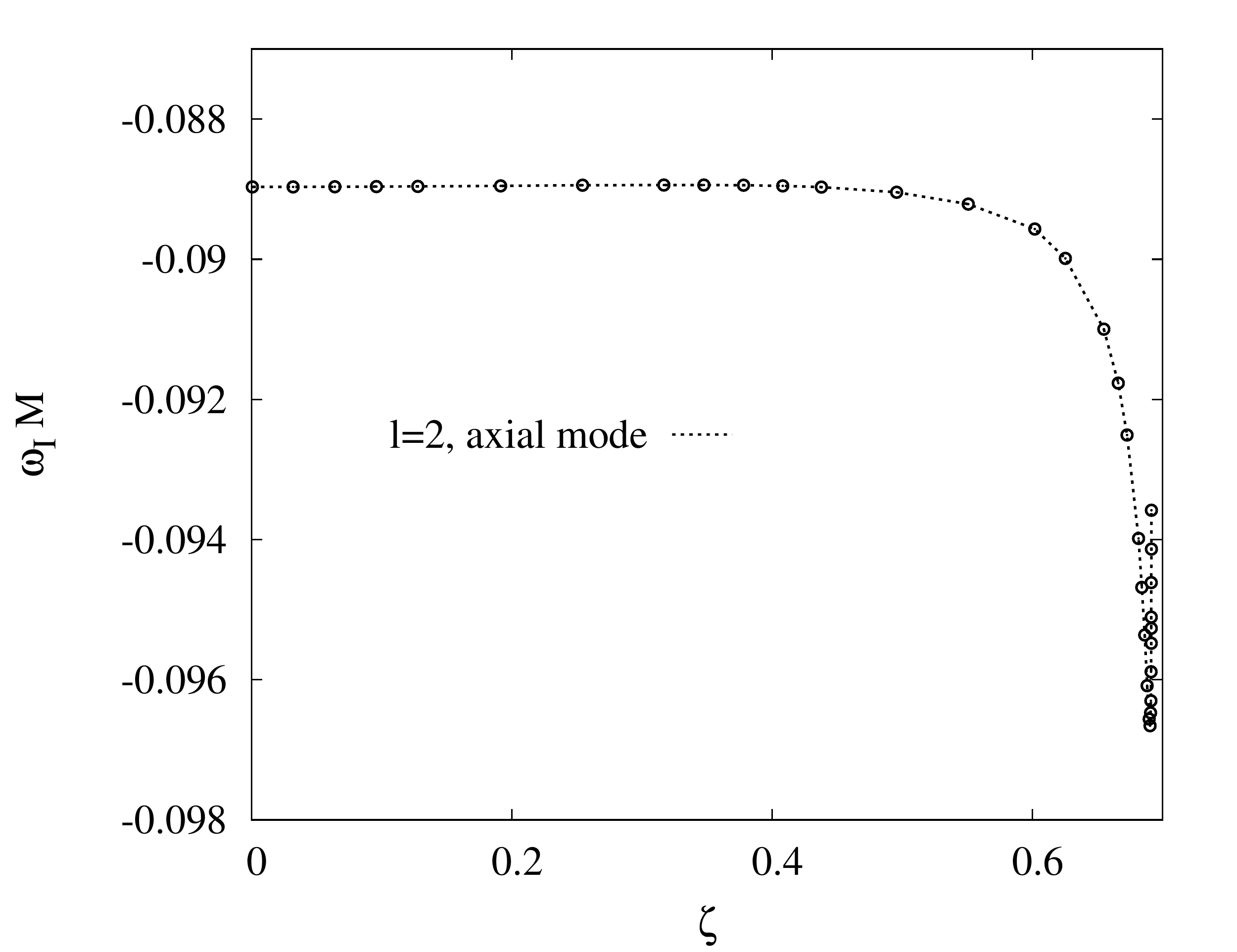}
\includegraphics[width=0.38\textwidth,angle=-90]{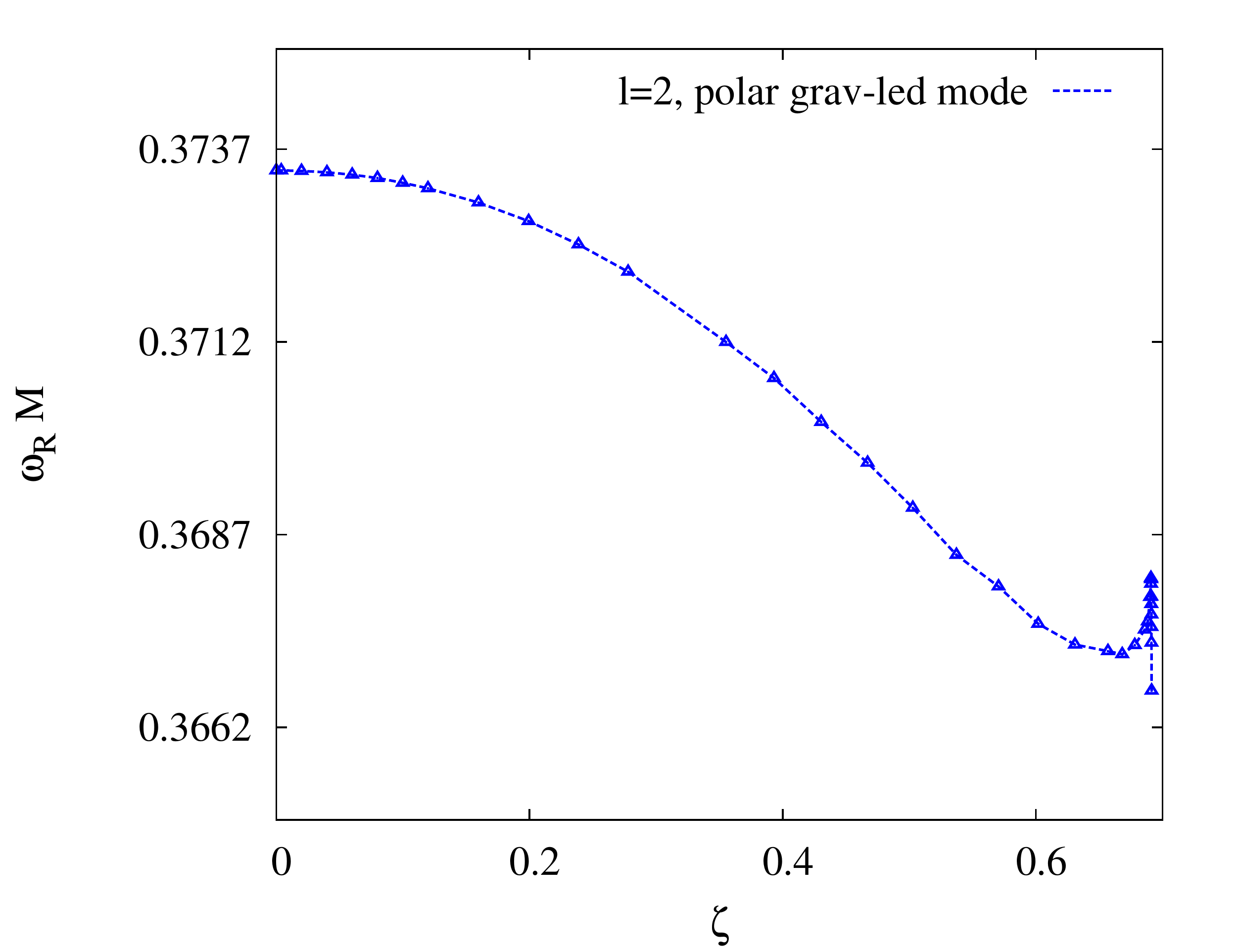}
\includegraphics[width=0.38\textwidth,angle=-90]{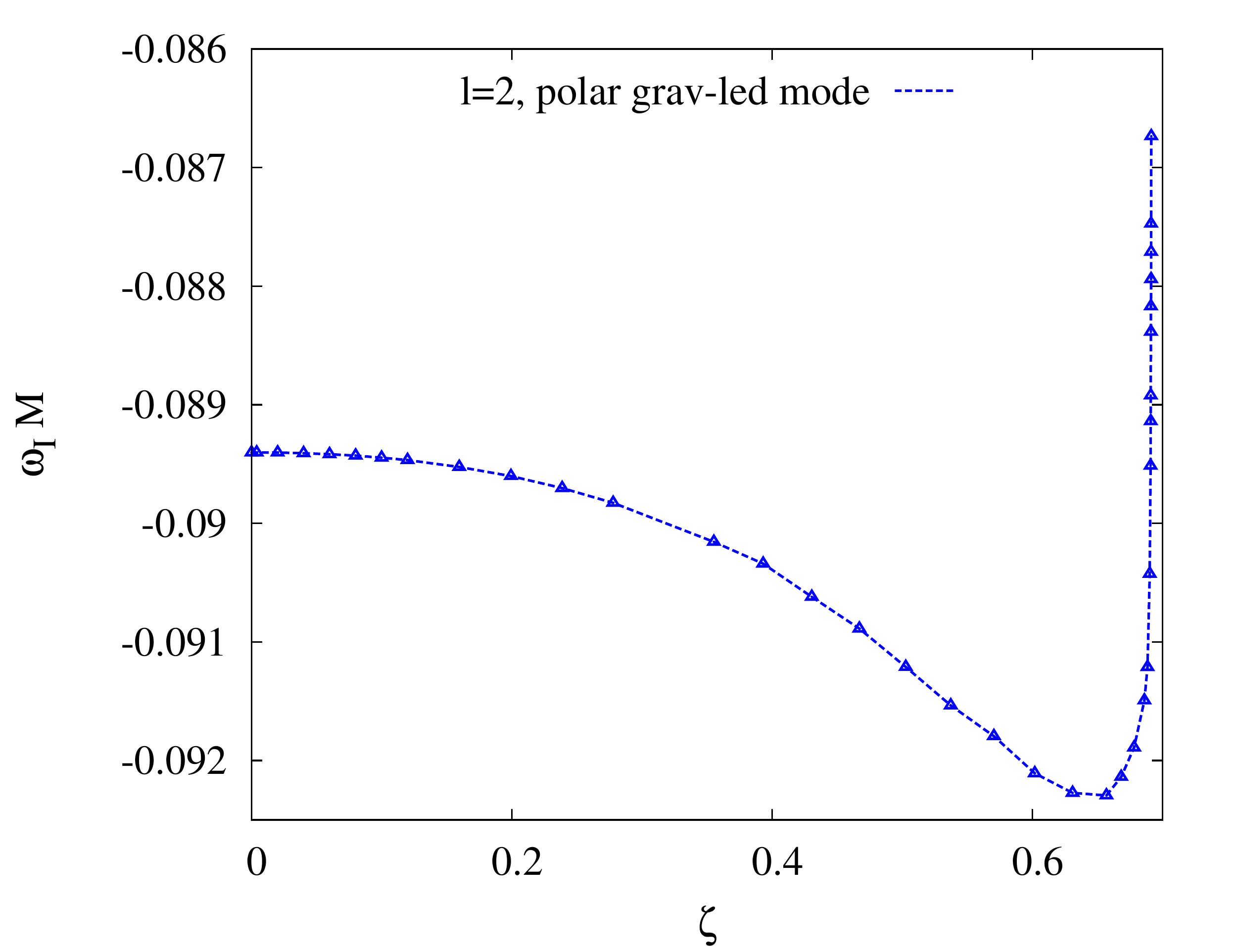}
\includegraphics[width=0.38\textwidth,angle=-90]{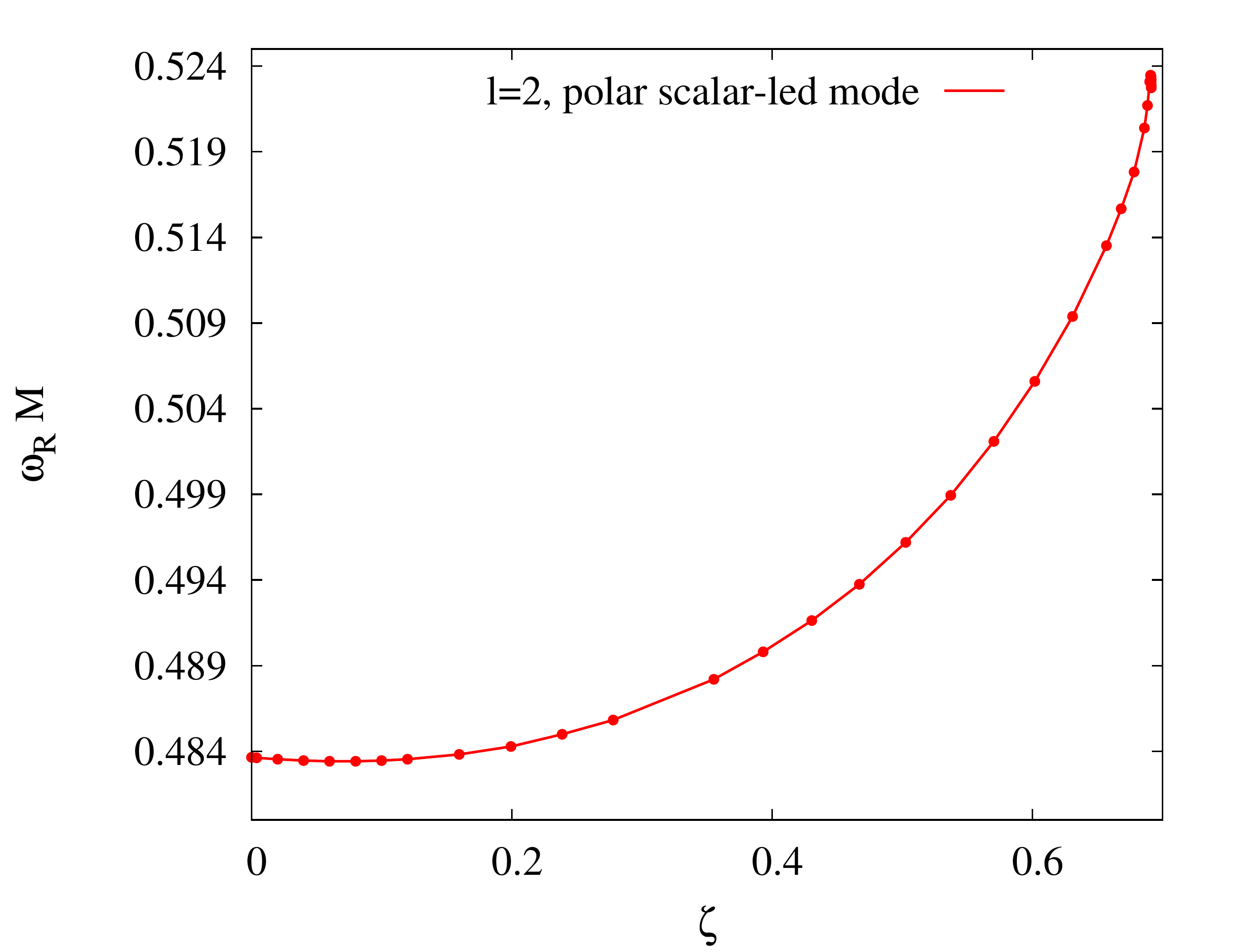}
\includegraphics[width=0.38\textwidth,angle=-90]{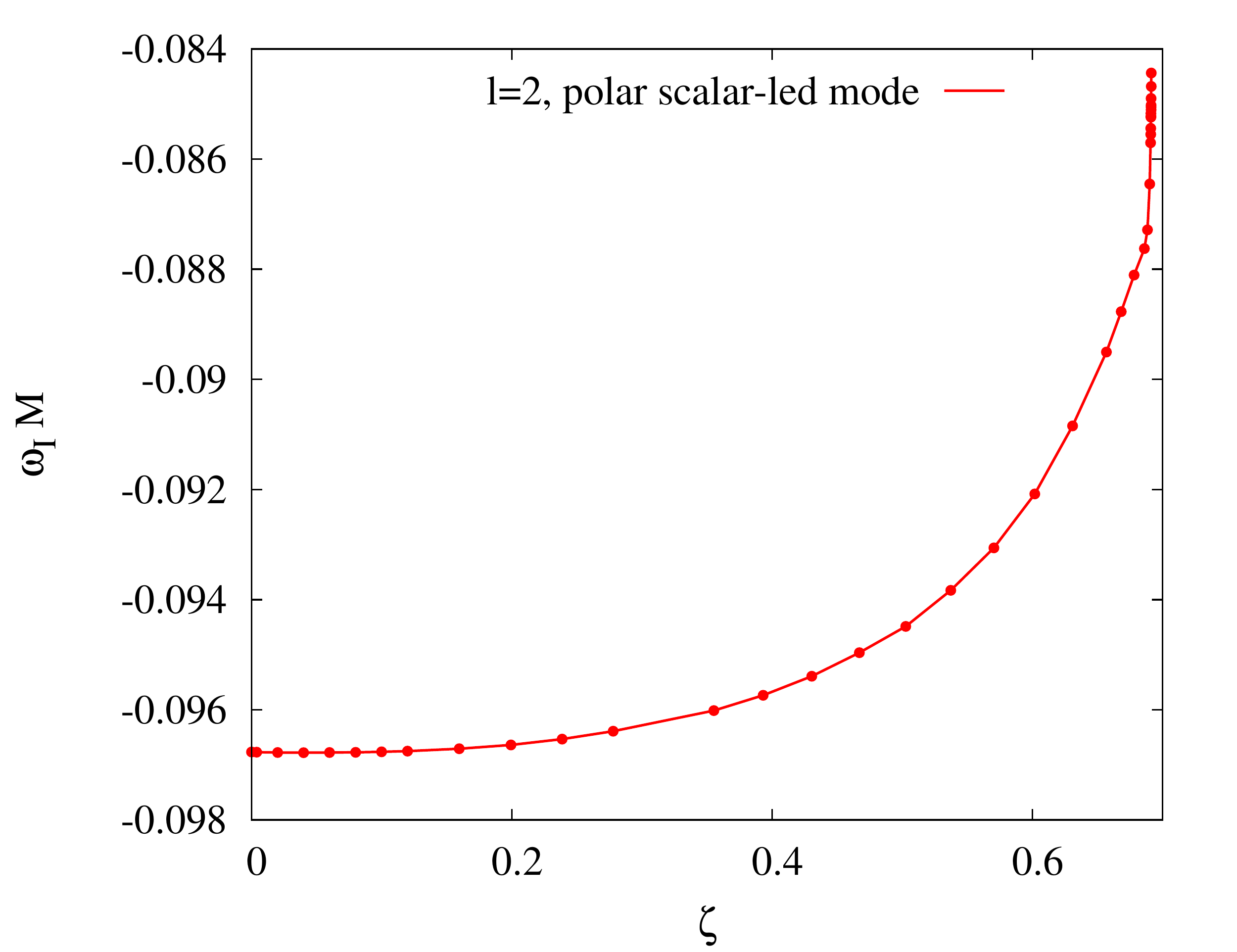}

    \caption{
The $l=2$ fundamental modes for type I dEGB black holes with $\gamma=1$. 
The left column shows the real part of the eigenvalue $\omega$ vs 
the scaled Gau\ss -Bonnet coupling constant $\zeta$, 
the right column the imaginary part of $\omega$. 
On top are the axial modes, in the middle the gravitational-led polar modes, 
and on the bottom the scalar-led polar modes.}
    \label{fig:dEGB_l2_modes}
\end{figure}

Let us now consider the spectrum of quasinormal modes 
for the type I black holes. 
In GR, the spectrum of quasinormal modes of the Schwarzschild solution 
is exactly the same in the polar and axial channels. 
This so-called isospectrality was first shown 
by Chandrasekhar \cite{Chandrasekhar:1985kt}, 
by relating analytically the polar and axial equations with a transformation.
For the dEGB black holes we are considering here,
this is no longer the case. 

For instance, consider the fundamental modes of the $l=2$ perturbations. 
We show the eigenvalues $\omega$ of these modes in Fig.~\ref{fig:dEGB_l2_modes} 
for the case $\gamma=1$. 
On the left we show the real part of $\omega$ vs the scaled coupling
constant $\zeta$, and on the right the imaginary part. 
The top set shows the axial modes, 
the set in the middle the gravitational-led polar modes, 
and the bottom set the scalar-led modes.
Here we use the term
gravitational-led modes, when the modes reduce to the
gravitational modes of Schwarzschild black holes in the limit $\zeta \to 0$,
while we use the term scalar-led modes, 
when the modes reduce to the quasinormal modes
of a test scalar field in a Schwarzschild metric in that limit.

In the limit $\zeta \to 0$, the gravitational-led polar modes 
coincide with the axial modes, since isospectrality is recovered. 
But as soon as the black holes have scalar hair, 
this isospectrality is broken, 
and axial modes differ from the gravitational-led polar modes. 
This is clearly seen in the discrepancies between the top set 
in Fig.~\ref{fig:dEGB_l2_modes} and the set in the middle. 
The frequency rises for the axial modes, while it drops for the polar modes, 
as the coupling $\zeta$ is switched on.
The damping times, which correspond to the
inverse of the imaginary part of the eigenvalue, differ as well
for $\zeta>0$.

However, not only isospectrality is broken, 
but also new modes appear in the spectrum. 
These new modes (shown in the bottom set of Fig.~\ref{fig:dEGB_l2_modes}) 
are related to scalar radiation. 
In the limit $\zeta \to 0$, 
these modes connect to the quasinormal modes of a test scalar field 
in the background of the Schwarzschild solution. 
When the coupling is finite, 
the scalar perturbations couple to the metric perturbations 
as seen in the previous section. 
Hence these modes also give rise to gravitational radiation.

These results imply that the spectrum of the resonant modes 
during the ringdown of a dEGB black hole is much richer 
than the spectrum of a simple Schwarzschild black hole in GR. 
The isospectrality of polar and axial modes is broken, 
and the presence of a scalar field sourced non-trivially by gravity 
introduces more resonances. 
Clearly, these effects are not exclusive of $l=2$ and $\gamma=1$, 
but they are present for arbitrary values of $l$ and $\gamma$. 

In this sense, the presence of a scalar field introduces 
new ways these hairy black holes can radiate. 
It is well known that in GR there is no monopolar or dipolar 
gravitational radiation -- hence only $l \geq 2$ modes appear in the ringdown. 
However, because of the scalar field these dEGB black holes 
also radiate monopolar $l=0$ and dipolar $l=1$ radiation, 
and these modes could appear naturally in the spectrum of resonances 
in the ringdown \cite{Blazquez-Salcedo:2017txk}.

In conclusion, the quasinormal mode spectrum of these dEGB black holes 
is much richer than the spectrum of Schwarzschild black holes, 
and these extra resonances could have some relevance 
in the observed spectrum of the ringdown phase of gravitational waves 
after the merger of black holes.

Let us now briefly discuss possible implications of this model in astrophysics. 
As commented above, the existence of a limiting $\zeta_L$ 
translates into a minimum value of the mass of dEGB black holes. 
It is possible to use this limiting value to obtain an upper constraint 
on the value of $\alpha$, 
using the smallest masses of black hole candidates observed in nature 
(by X-rays or gravitational waves). 
Since $\zeta_L$ depends on the dilaton coupling $\gamma$,
so does the constraint \cite{Blazquez-Salcedo:2017txk}.

Concerning the modes, we have seen that the largest deviation 
with respect to the Schwarzschild spectrum arises 
when the value of $\zeta_L$ is approached 
(i.e., the minimum value of the mass). 
Hence, we expect that when using observations of the ringdown phase 
in order to (further) constrain the value of the coupling $\alpha$, 
it will be more effective to observe the ringdown phase 
of small black holes (for example after mergers of neutron stars, 
which are expected to be closer to $2 M_{\odot}$). 
The ringdown of large dEGB black holes will look very similar 
to the ringdown of Schwarzschild black holes, 
and the frequencies could be very difficult to distinguish. 
However, a smoking gun could be the detection of 
monopolar/dipolar radiation and extra scalar modes in the ringdown phase.

\subsection{Scalarized solutions} \label{sec_sEGB}

Let us discuss next a new type of black holes with scalar fields,
discovered recently in scalar-Einstein-Gau\ss -Bonnet (sEGB) theory,
also referred to as extended scalar-tensor-Gau\ss -Bonnet (estGB) theory
\cite{Doneva:2017bvd,Silva:2017uqg,Antoniou:2017acq,Antoniou:2017hxj,Blazquez-Salcedo:2018jnn}. 
If the coupling function $f(\phi)$ is chosen to have a quadratic leading term 
of the form $f(\phi) \sim \phi^2$, 
then the theory features spontaneous scalarization 
of the Schwarzschild solution. 
This means that scalarized black holes arise for a certain
range of the coupling constant, 
while the Schwarzschild black hole is still a solution of the theory 
for arbitrary values of the coupling constant.
This is clearly different from the dEGB case, 
where the Schwarzschild black hole is no longer a solution
of the field equations. In dEGB theory, the Schwarzschild solution is only approached asymptotically 
for large values of the black hole mass. 
In estGB theory, branches of scalarized black holes bifurcate 
from a discrete set of Schwarzschild black holes 
at critical values of the mass. 
We note, that such scalarization of black holes has also been found 
in other theories (see e.g., \cite{Stefanov:2008,Doneva:2010,Herdeiro:2018wub}).

In this section we will focus on a particular exponential coupling function
\begin{equation}
f(\phi)=  \frac{1}{12} \left(1- e^{-6\phi^2}\right). \label{eq:coupling_function}
\end{equation}
This coupling function captures many of the general properties 
of other coupling functions such as, 
in particular, the spontaneous scalarization sourced by ${\cal R}^2_{GB}$, 
and the existence of multiple branches of black holes. 
However, this coupling function presents some especially interesting features,
and it mimics the coupling function usually considered 
in spontaneus scalarization of neutron stars in STTs.

\begin{figure}[t]
     \centering
\includegraphics[width=0.38\textwidth,angle=-90]{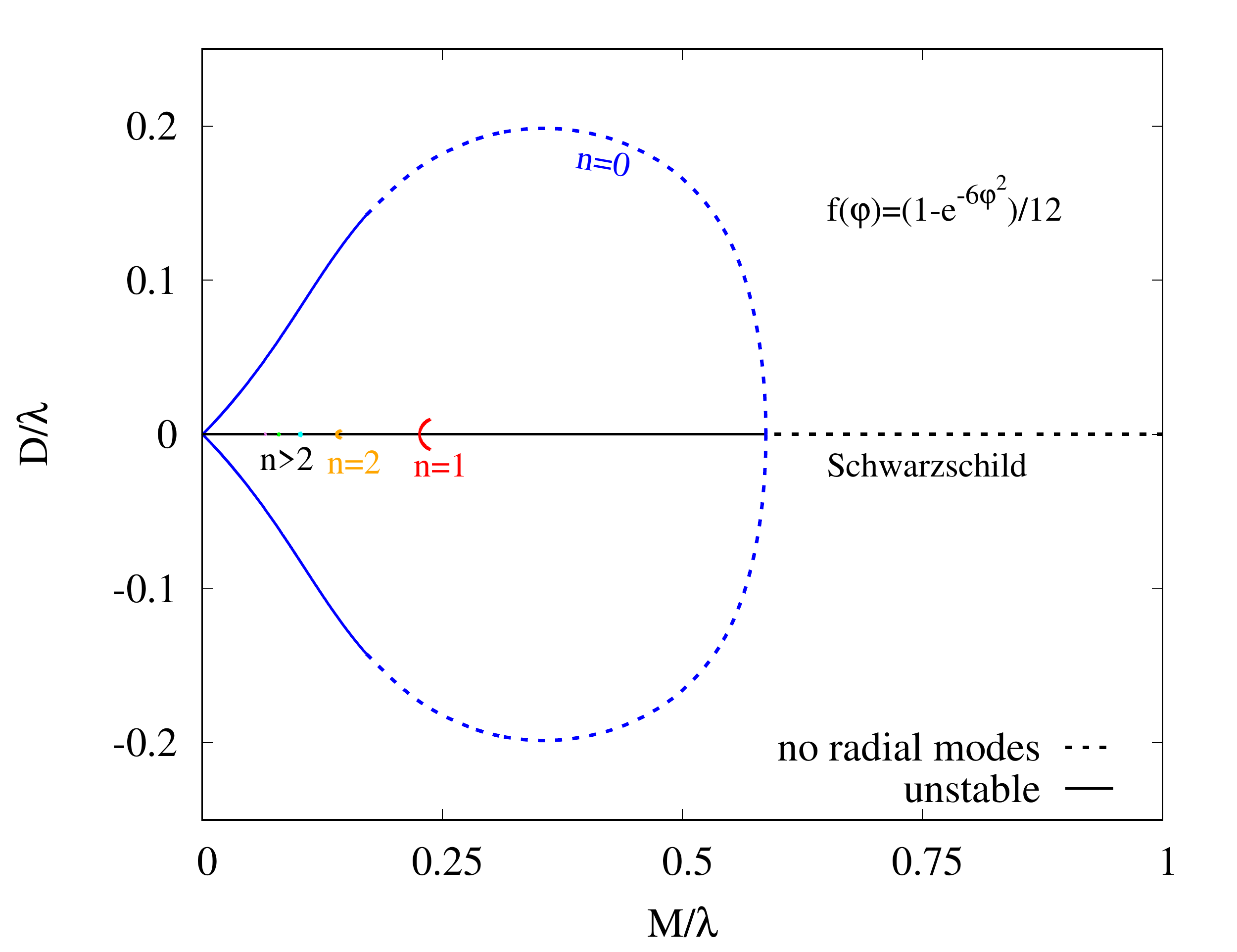}
\includegraphics[width=0.38\textwidth,angle=-90]{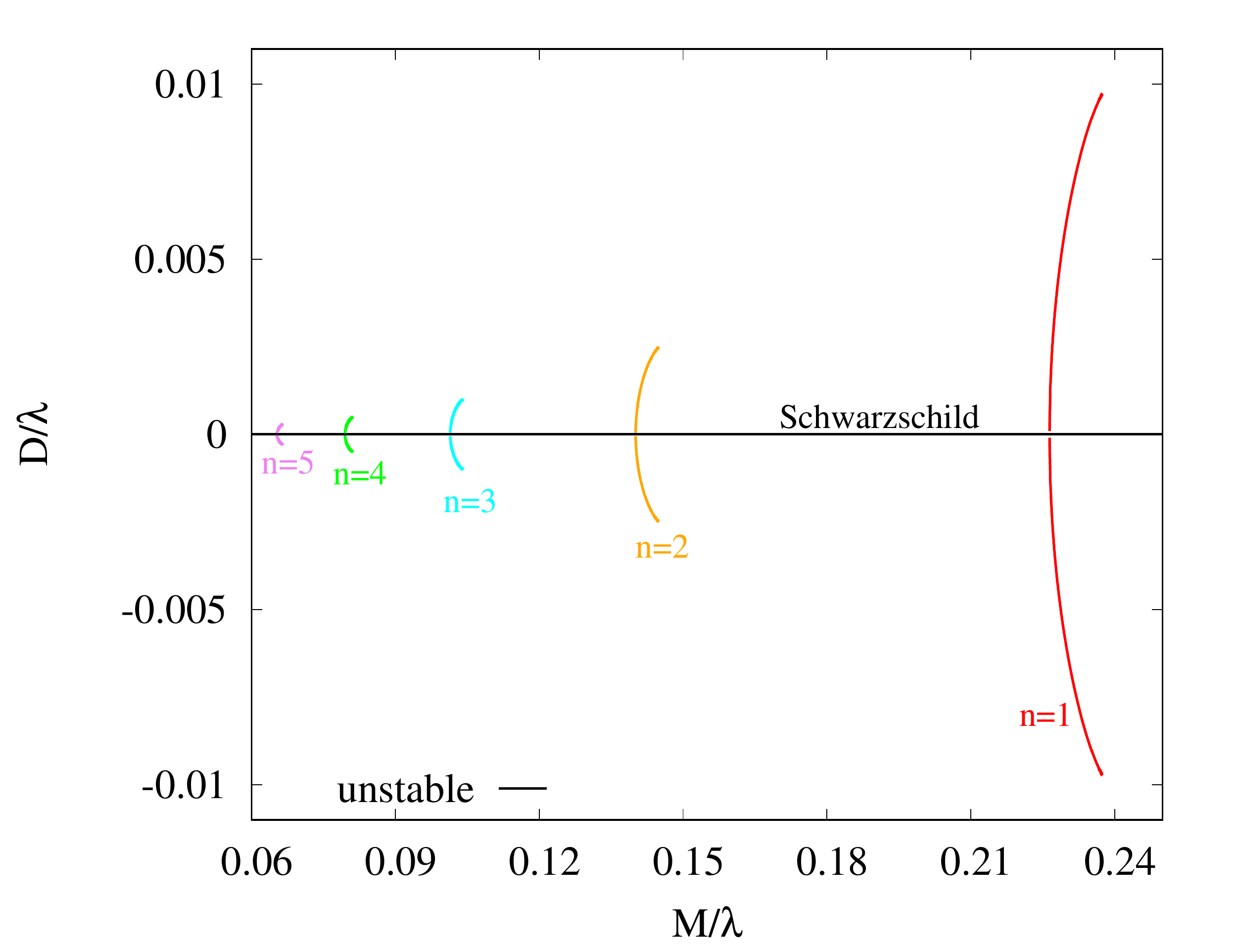}

        \caption{The scalar charge $D$ vs the mass $M$ 
(both scaled with $\lambda$) of scalarized black holes. 
The left panel shows the Schwarzschild solution (black),
and the scalarized branches with $n=0$ (blue), $n=1$ (red), 
$n=2$ (orange) as well as some very small $n>2$ branches. 
The right panel is a zoom to show more clearly the 
$n=3$ (cyan), $n=4$ (green) and $n=5$ (purple) branches.
Dashed lines indicate that the solutions have no unstable modes.}
         \label{fig:sEGB_MD}
\end{figure}

In Fig.~\ref{fig:sEGB_MD} we show the scalar charge $D$ 
vs the mass $M$ (scaled with $\lambda$), 
which summarizes the space of solutions for this coupling. 
The Schwarzschild branch of solutions is shown in black. 
At some critical values of $M/\lambda$, 
branches of scalarized black holes bifurcate 
from the Schwarzschild solution. 
Numbering these branch with an integer number $n$, 
this number is related to the number of nodes 
present in the scalar field $\phi$, as it extends from the horizon to infinity. 
In Fig.~\ref{fig:sEGB_MD} the $n=0$ branch is depicted in blue. 
Interestingly, it bends back to $M/\lambda=0$. 
The other scalarized branches 
($n=1$ in red, $n=2$ in orange, $n=3$ in cyan, $n=4$ in green and $n=5$ in purple) 
extend from the bifurcation point with the Schwarzschild solution 
up to a critical solution, where the condition (\ref{eq:BC_sqrt_rh}) 
is no longer satisfied.

Following the same steps as for the dilatonic black holes, 
one can study the quasinormal mode spectrum of the scalarized black holes. 
However, so far only radial perturbations have been considered 
\cite{Blazquez-Salcedo:2018jnn}. 
Interestingly, some simple couplings 
do not allow for the appearence of stable scalarized configurations.
(This is, for instance, the case for the simplest quadratic coupling.)
Clearly,
not all scalarized black hole branches may actually be astrophysically relevant.

\begin{figure}[t]
     \centering
\includegraphics[width=0.38\textwidth,angle=-90]{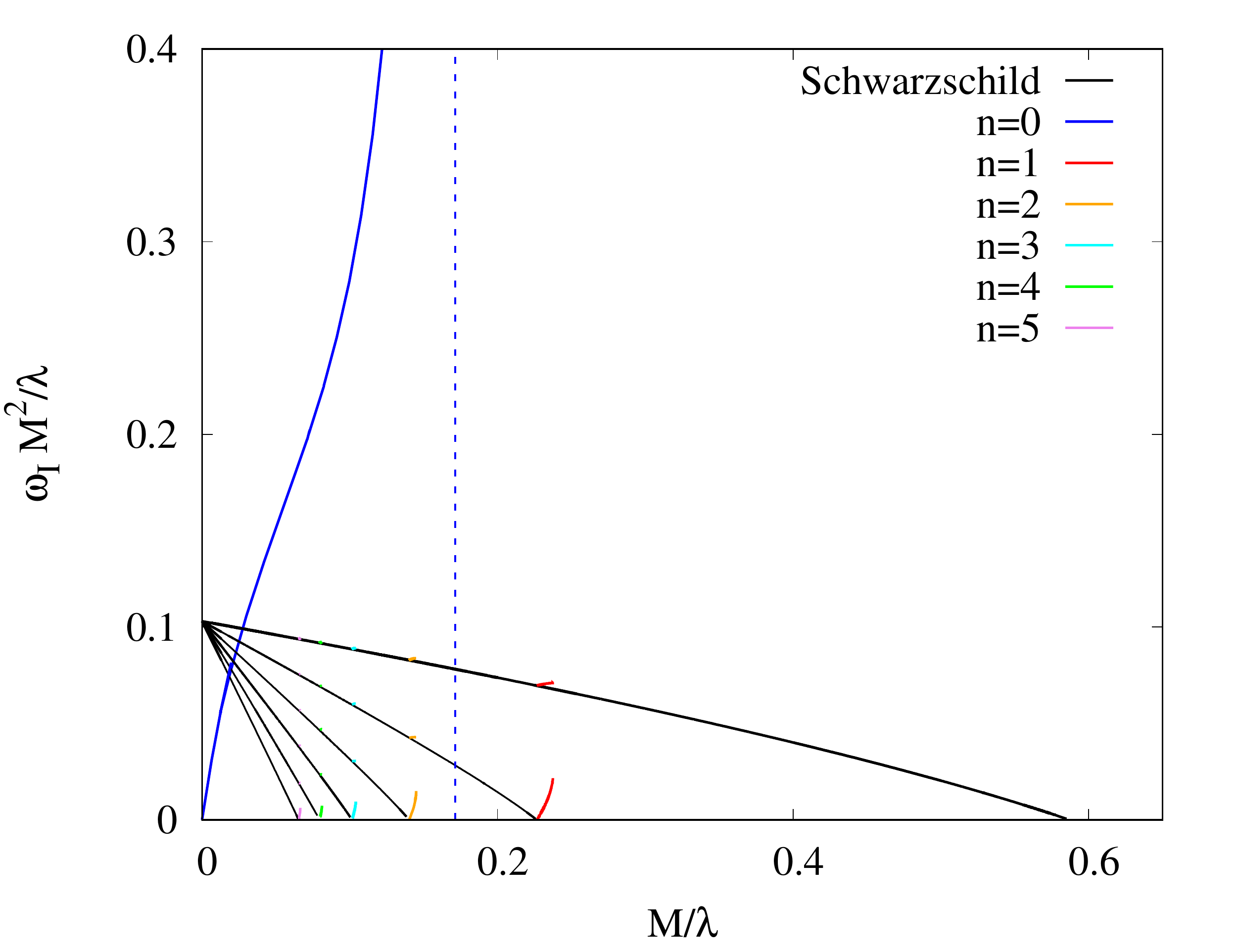}
\includegraphics[width=0.38\textwidth,angle=-90]{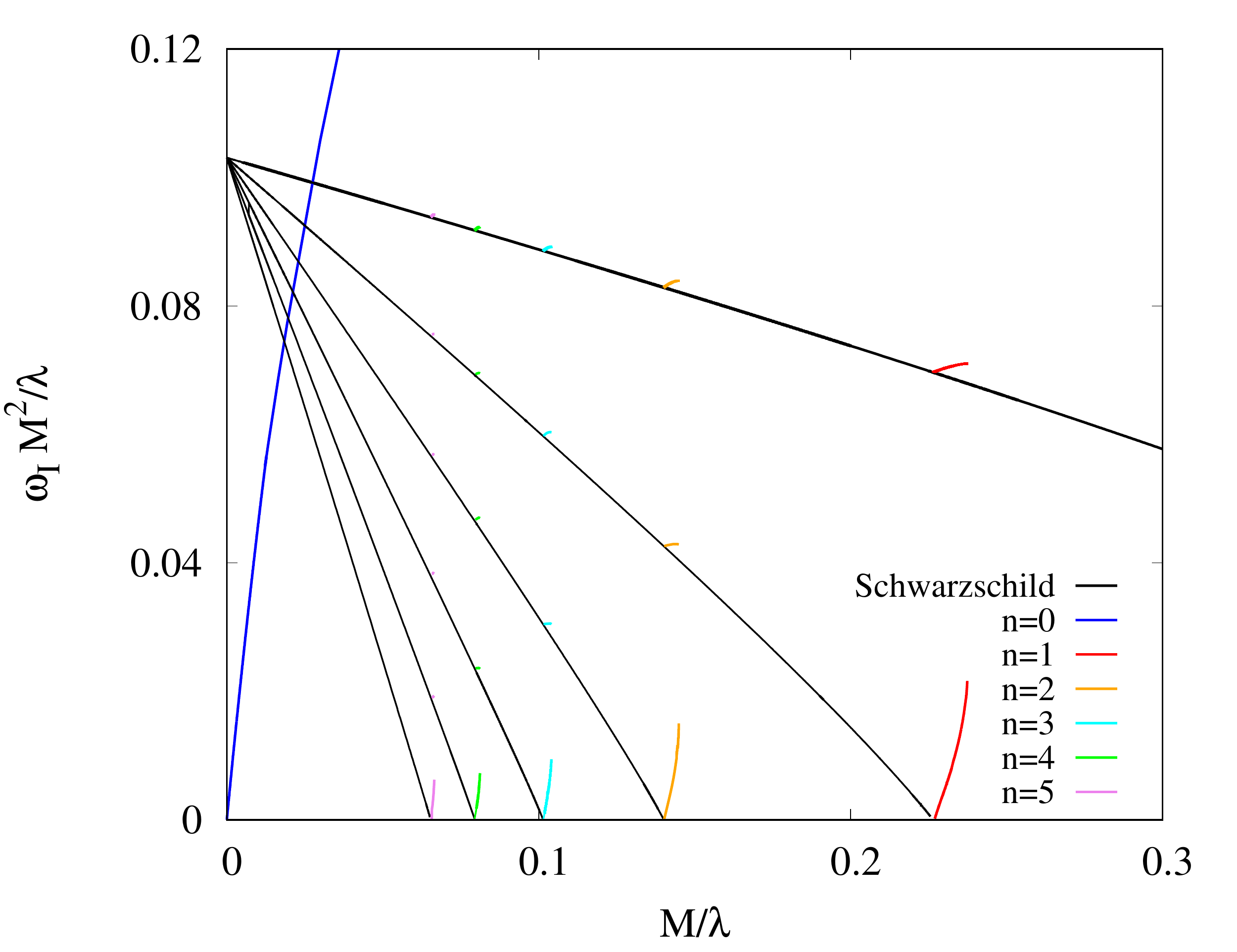}

         \caption{(left) Scaled imaginary part $\omega_I M^2 /\lambda$ 
of the $l=0$ unstable modes vs the scaled mass $M/\lambda$
of scalarized black holes.
(right) A zoom into the region of parameter space 
with the unstable modes of the scalarized branches with higher $n$.}
         \label{fig:sEGB_unstable}
\end{figure}

In Fig.~\ref{fig:sEGB_unstable} we show the unstable $l=0$ modes 
for the black hole solutions with coupling (\ref{eq:coupling_function}). 
We follow the same color coding as in Fig.~\ref{fig:sEGB_MD}, 
with the modes of the Schwarzschild solution in black, $n=0$ in blue, etc. 
Indeed, the figure shows numerous unstable modes.
For instance, the branch of Schwarzschild black holes 
becomes unstable for $M/\lambda<0.587$, 
and the number of unstable modes grows, as $M/\lambda$ decreases,
whenever a zero mode 
associated with the bifurcation of a new branch 
of scalarized black holes is passed.

The scalarized branches with $n>0$ are also unstable, 
with unstable modes bifurcating from the Schwarzschild zero mode, 
but also from the unstable modes of the Schwarzschild solution, 
forming together a tower of modes.
For instance, the $n$th branch possesses $n+1$ unstable modes. 
The $n=0$ branch is special, though, 
because it does not show any unstable mode bifurcating 
from the first zero mode. 
However, it possesses an unstable mode in the region 
of parameter space $M/\lambda < 0.171$. 
This mode is related to the loss of hyperbolicity 
of the perturbation equations and it will require a more detailed study. 
Nonetheless, the $n=0$ branch of scalarized black holes 
is free of instabilities, when $M/\lambda > 0.171$. 
This is in contrast to the case of the quadratic coupling, 
where this $n=0$ branch is also mode unstable.

From the analysis of the stability of these scalarized black holes 
we learn that, for a fixed value of the coupling $\lambda$, 
there exists also an effective minimum value of the mass. 
The minimum value of the hairless Schwarzschild solution 
is $M_{\min}=0.587\lambda$, 
and this can be used to constrain the coupling constant 
using the values of the smallest black hole candidates observed in nature. 
The spectrum of quasinormal modes of these black holes 
is the one predicted by GR, 
together with the scalar modes of a minimally coupled scalar field. 

On the other hand, the effective minimum mass of a scalarized black hole 
is smaller, with $M_{\min}=0.171\lambda$. 
Because of the presence of a non-trivial scalar field, 
the quasinormal modes that dominate the ringdown phase 
of these scalarized black holes will be different from Schwarzschild, 
and they will be qualitatively similar 
to the ones of the dilatonic black holes: 
they will feature broken isospectrality, 
and the spectrum will also have a contribution from scalar-led modes, 
including monopolar and dipolar radiation. 
In this case, scalarized black holes can only exist 
for a limited range of masses, 
and the influence of the scalar field on the quasinormal mode spectrum 
is expected to be larger for small scalarized black holes 
close to the limit $M_{\min}=0.171\lambda$. 
However, a more detailed investigation needs to be done, 
and will be presented elsewhere.

\section{Neutron stars with scalar fields} \label{sec_ns}

In this section we will discuss the spectrum of quasinormal modes 
of neutron stars in several alternative theories of gravity,
that also include a scalar field or can be reformulated
in terms of a scalar field.
In particular, we address neutron stars in
$R^2$ gravity, STT, Horndeski gravity, and dEGB theory.

Like black holes, neutron stars are compact objects,
allowing testing of alternative theories of gravity.
But unlike black holes,
which do not need any extra matter to be sustained,
neutron stars are much harder to model than black holes, 
since they are composed of nuclear matter in a very dense state. 
As discussed above, the neutron star composition 
can be described in terms of an EOS,
relating the energy density of the fluid to the pressure. 
However, the EOS is not well known at very high densities, 
and numerous models for the nuclear EOS have been proposed.

This fact complicates the analysis of the quasinormal mode spectrum, 
since one has to take into account the indeterminacy
introduced in the modes by the lack of knowledge of the proper EOS. 
In practice, all one can do under these circumstances
is to calculate the spectrum for a large variety of EOSs,
that have not yet been excluded by observations.
In the following we will discuss that the indeterminacy 
introduced by the EOSs on the mode spectrum
can, in principle, compete with the indeterminacy 
introduced by employing different gravity theories
with their currently allowed ranges of coupling constants.

Under these circumstances universal relations may prove to be of great help
\cite{Yagi:2016bkt,Doneva:2017jop}.
Universal relations between global quantities of neutron stars,
like the $I$-Love-$Q$ relations, exhibit for the properly scaled quantities
a remarkable EOS independence.
In particular, realistic neutron star models 
satisfy universal relations that are largely matter independent, 
i.e., these relations are satisfied with rather good
accuracy for numerous realistic EOSs.
Likewise, there are universal relations
between the quasinormal modes and the global quantities of the neutron stars. 
Such universal relations have been widely studied in GR. 

In the next subsections we will address such universal relations 
for different theories of gravity, 
and discuss how these relations could be used 
to extract information about the theory of gravity 
from observations of the ringdown of a neutron star. 
{Our main focus will be on the axial modes of neutron 
stars in these alternative theories of gravity. The 
polar modes will be briefly reviewed since the 
problem of the calculation of the polar modes in 
alternative theories of gravity is not fully solved, and
most of the studies employ
the Cowling approximation, where the spacetime and scalar degrees of 
freedom are neglected.}
We will start with a special $f(R)$ gravity, called $R^2$ gravity
\cite{Sotiriou:2008rp,DeFelice:2010aj,Capozziello:2009nq,Capozziello:2011et}.

\subsection{$R^2$ gravity} \label{sec_R2}

In this subsection we discuss the axial perturbations of neutron stars 
in $f(R)$ gravity with Lagrangian $f(R) = R + \tilde a R^2$. 
However, instead of using the $f(R)$ action and field equations, 
we will analyze the equivalent STT in the Einstein frame. 
A detailed discussion of the connection between the $f(R)$ theory
and the STT in the Jordan and Einstein frames 
can be found in \cite{Yazadjiev:2014cza, Staykov:2014mwa, Yazadjiev:2015zia}.

The action of the theory in the Einstein frame is given by
\begin{eqnarray} \label{eq:action_R2}
S = \frac{1}{16\pi G}\int d^4x\sqrt{-g}
\Big[R - 2\nabla_\mu \phi \nabla^\mu \phi  - V(\phi) 
+ L_{\rm matter}(A^2(\phi) g_{\mu\nu},\xi)\Big] ,
\end{eqnarray}
with the coupling function $A(\phi)$ and the potential $V(\phi)$,
\begin{eqnarray}\label{eq:CouplingFunc_R2}
A(\phi)=e^{-\frac{1}{\sqrt{3}}\phi}, \;\; \; 
V(\phi)= \frac{1}{4\tilde a} \left(1-e^{-\frac{2\phi}{\sqrt{3}}}\right)^2.
\end{eqnarray}
Note, that the coupling function $A(\phi)$ enters into 
the matter Lagrangian $L_{\rm matter}$. 
This means, that the perfect fluid description 
and the equation of state have to be calculated in the Jordan frame, 
which represents the physical frame,
and then transformed with the help of $A(\phi)$ into the Einstein frame. 
The explicit transformation is found in 
\cite{Yazadjiev:2014cza, Staykov:2014mwa, Yazadjiev:2015zia}.

In this theory the scalar field possesses a mass 
determined by the coupling parameter $\tilde a$, 
$m_\phi=1/\sqrt{6 \tilde a}$. 
The limit $\tilde a\rightarrow \infty$ then leads to $m_\phi=0$, 
which corresponds to a particular subclass of massless 
Brans-Dicke theories. 
In the limit $\tilde a\rightarrow 0$ 
the mass of the scalar field tends to infinity, 
and the theory tends to GR.

\begin{figure}[h]
	\centering
	\includegraphics[width=0.4\textwidth,angle=-90]{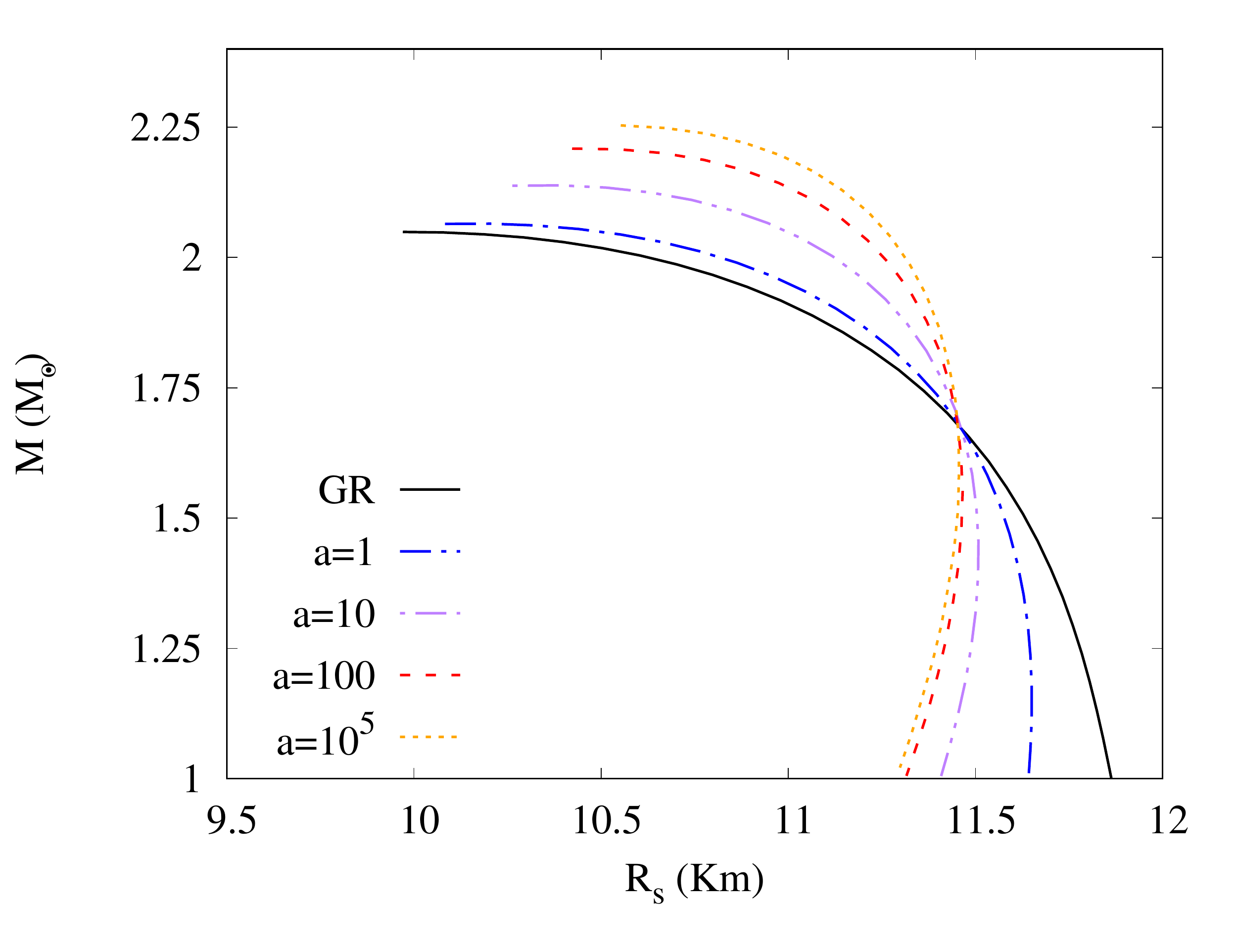}

	\caption{Mass-radius relation for neutron stars in $R^2$ gravity
with EOS SLy for several values of the coupling $a$;
black solid line: GR, blue dash-dotted line: $a=1$, 
purple dash-dot-dot line: $a=10$, red dashed line: $a=100$,
orange dotted line: $a=10^5$.}
	\label{R2_MR}
\end{figure}

Based on the Ansatz for spherically symmetric neutron stars,
Eqs.~(\ref{ds2_0})-(\ref{phi0}), one first obtains numerically 
the configurations for various EOSs. 
In $R^2$ gravity in the Einstein frame
the neutron stars possess a non-trivial scalar field 
that extends from the center of the star up to infinity. 
However, when the field is massive, it decays exponentially towards infinity. 
The more massive the field the more damped it is. 
The properties of neutron stars in $R^2$ gravity have been studied in 
\cite{Yazadjiev:2014cza,Staykov:2014mwa,Yazadjiev:2015zia,Astashenok:2017dpo}.

Typically, the resulting neutron star models are not very 
different from their GR counterparts. 
This is demonstrated in Fig.~\ref{R2_MR},
where we show as an example the mass-radius relation 
for a particular EOS (here SLy) and several values of the
coupling constant $a$. 
The larger the value of $a$, 
the stronger the scalar field grows, 
and the more the mass-radius relation changes as compared to GR.

\begin{figure}[h]
	\centering
	\includegraphics[width=0.37\textwidth,angle=-90]{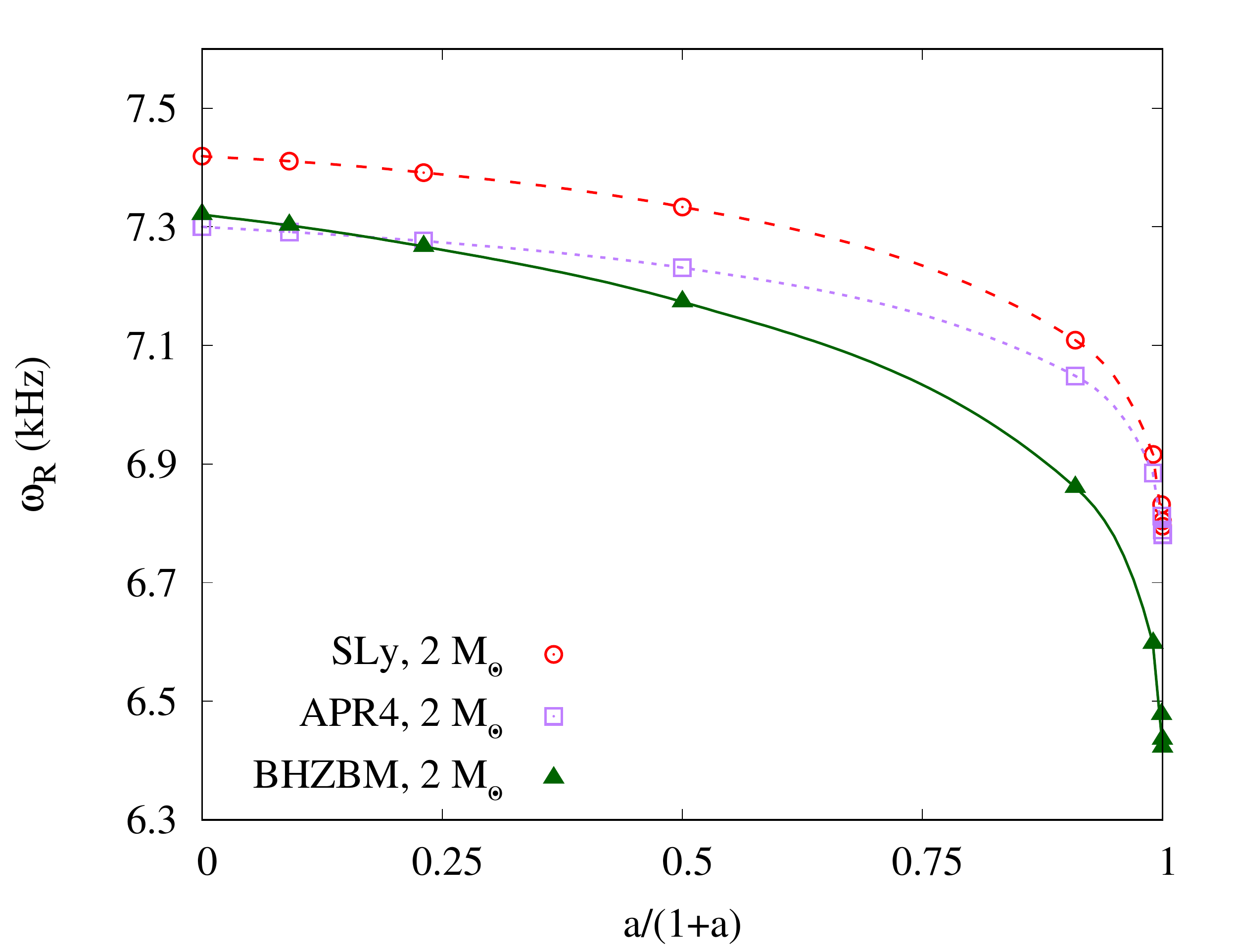}\label{l2_2M_real}
	\includegraphics[width=0.37\textwidth,angle=-90]{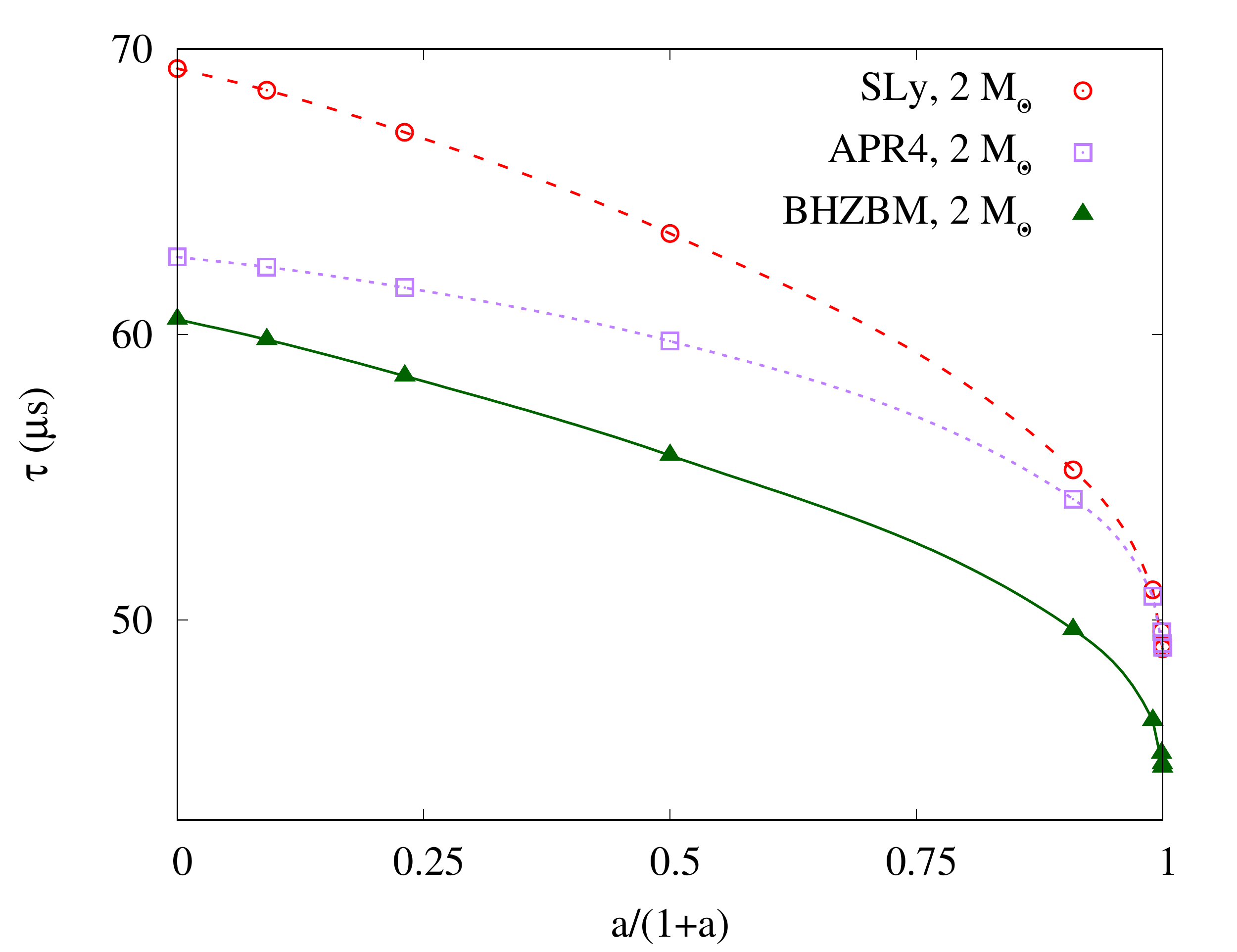}\label{l2_2M_img}
	\caption{(left) The frequency $\omega_R$ in kHz vs the parameter $a$
(compactified) for the fundamental $l=2$ curvature mode 
of neutron stars with a mass of 2 $M_\odot$ in $R^2$ gravity. 
The colors indicate the EOSs: Sly in brown with crosses, 
APR4 in green with squares, BHZBM in purple with circles.
(right) The analogous figure for the damping time $\tau$ in $\rm \mu s$.}
	\label{l2_2M}
\end{figure}

Let us now turn to the quasinormal mode spectrum for axial perturbations, 
first studied in \cite{Blazquez-Salcedo:2018qyy}. 
In Fig.~\ref{l2_2M} we show the frequency (left) and the damping time (right) 
versus the parameter $a=\tilde{a}/(1.49 \ \text{km})$,
where we fix the mass of the neutron star to 2 $M_\odot$, 
and present the modes for three realistic EOSs: 
Sly in brown with crosses, APR4 in green with squares, 
BHZBM in purple with circles. 

When $a \to 0$ the configurations tend to the pure GR solution, 
and when $a \to \infty$, to a massless Brans-Dicke theory, 
for which the deviation from GR becomes maximal. 
However, the figure shows, that the same value of %
the frequency or the damping time
can be obtained for different combinations of 
the parameter $a$ and the EOS. 
For instance, in the left panel 
we can get any frequency between $6.8$ kHz and $7.3$ kHz 
by tuning the EOS and the parameter $a$.
In the right panel we can get any damping time 
between $50$ $\mu$s and $60$ $\mu$s. 
Including more EOSs makes the degeneracy only worse. 
Hence a priori we cannot distinguish the effect 
of the gravitational theory (subject to the choice of value of $a$), 
from the effect of the properties of the high density matter 
(subject to the choice of the EOS).

\begin{figure}[h]
	\centering
	\includegraphics[width=0.49\textwidth,angle=-90]{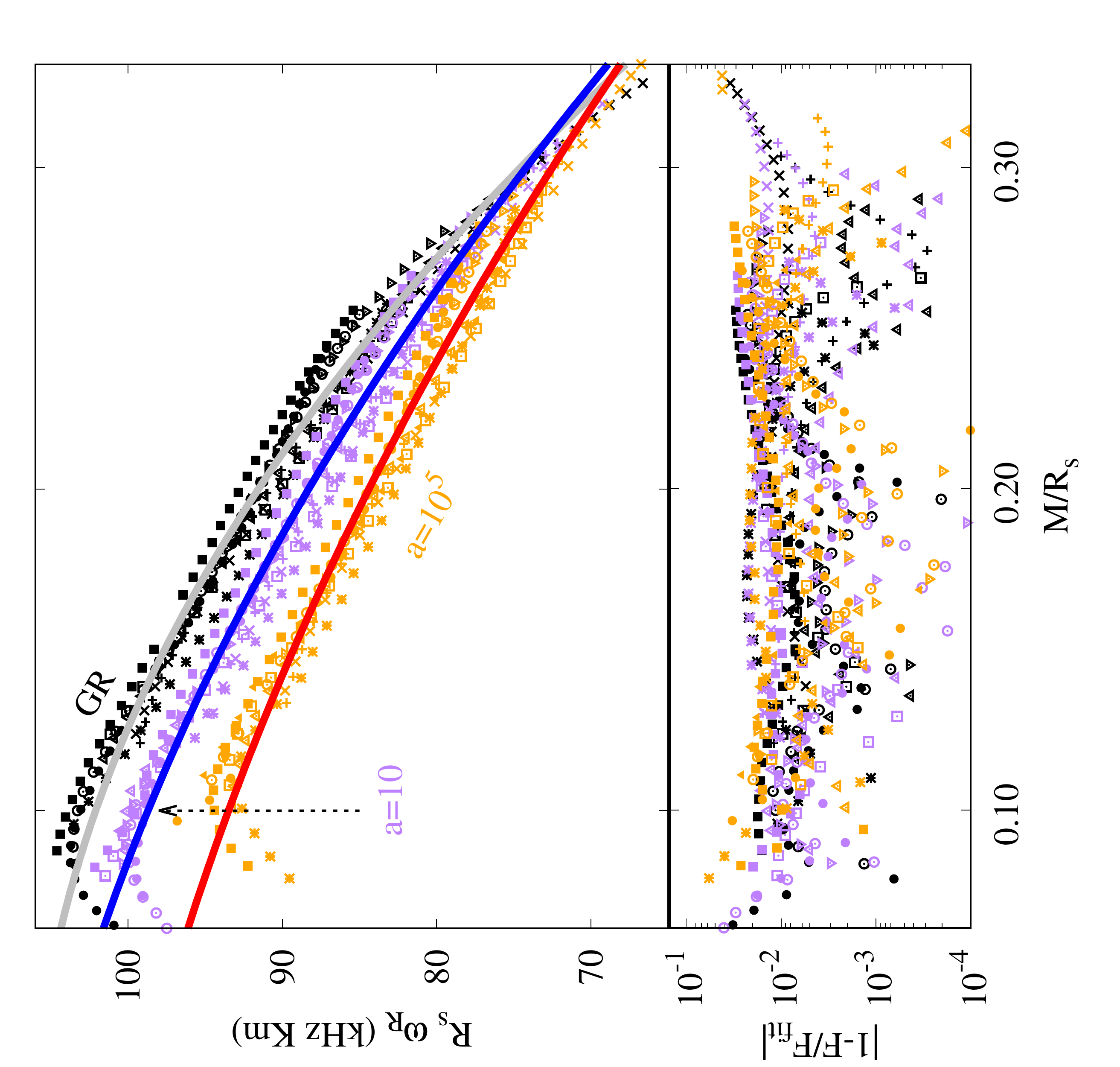}
	\includegraphics[width=0.49\textwidth,angle=-90]{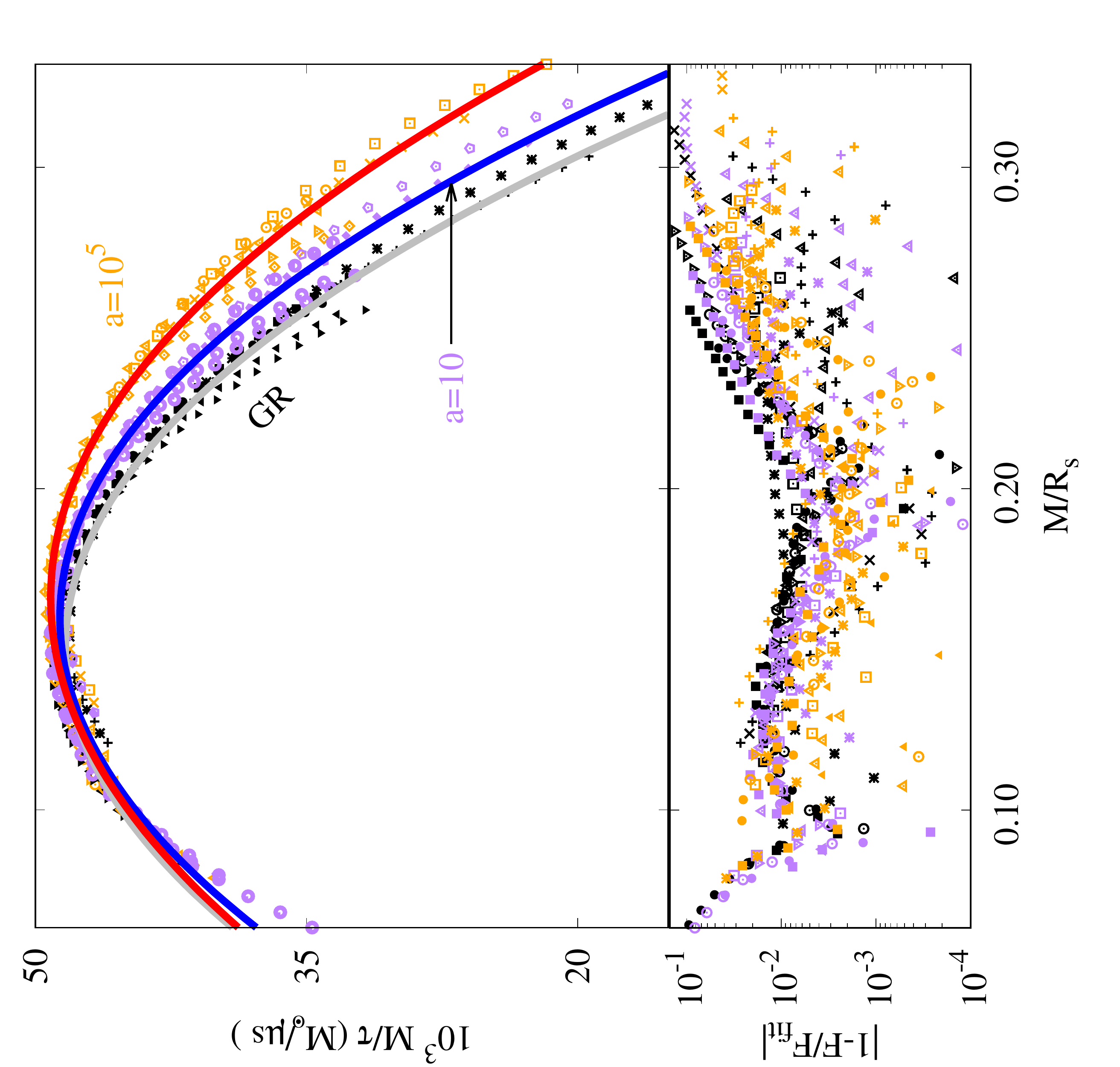}
	\caption{(left) The frequency $\omega_R$ scaled by the radius 
$R_s$ of the star (in $\rm kHz \cdot km$) vs the compactness $M/R_s$ 
for neutron stars in $R^2$ gravity.
(right) Inverse of the damping time $\tau$ scaled 
by the mass (in $M_{\odot}/\mu \rm s$) 
vs $M/R_s$. Ten different EOSs are shown (see text) for three 
values of the coupling $a$,
using black for GR, purple for $a=10$, and orange for $a=10^5$.
The solid lines correspond to quadratic fits 
(grey for GR, blue for $a=10$ and red for $a=10^5$).
In the bottom panels the difference between the fits and the data is shown, 
with $F$ denoting the corresponding scaled quantity in the panel above.
}
	\label{R2_scaled}
\end{figure}


In order to (almost) remove the matter dependence, 
we now construct scaled relations 
using the spectrum and the global quantities of the stars 
such as their total mass and their radius $R_s$ 
\cite{Andersson:1996pn,Andersson:1997rn,Kokkotas:1999mn,Benhar:2004xg,Blazquez-Salcedo:2015ets}.
These relations then constitute a set of universal relations
for the neutron stars.
In Fig.~\ref{R2_scaled} we show the scaled spectrum 
versus the compactness $M/R_s$ of the neutron stars.
The left panel exhibits the frequency $\omega_R$ 
scaled by the radius $R_s$ 
and the right panel the damping time $\tau$ scaled by the mass. 

The figure contains the data for ten different EOSs containing
purely nuclear matter (SLy \cite{Douchin:2001sv} , APR4 \cite{Akmal:1998cf}), 
hyperon matter (BHZBM \cite{Bednarek:2011gd}, GNH3 \cite{Glendenning:1984jr}, 
H4 \cite{Lackey:2005tk}, WCS1-2 \cite{Weissenborn:2011ut}), 
and a mixture of quark and nuclear matter 
(ALF2-4 \cite{Alford:2004pf}, WSHPS3 \cite{Weissenborn:2011qu}).
Note, that for the EOSs we follow the nomenclature introduced in 
\cite{Blazquez-Salcedo:2015ets,BlazquezSalcedo:2012pd,Blazquez-Salcedo:2013jka,
Blazquez-Salcedo:2018qyy,Motahar:2017blm}. 

Moreover, the data is divided in three sets for three values of the
coupling constant $a$, shown
in black for GR, in purple for $a=10$ and in orange for $a=10^5$. 
Each of these sets is fitted by a certain quadratic relation of the form 
\begin{eqnarray} 
\label{fits_R2a}
\omega_R[\rm kHz]\cdot R_s [\rm km] = 
a_0 + a_1 \left( \frac{M}{R_s} \right) + a_2 \left( \frac{M}{R_s} \right) ^2 \ ,\\
\frac{M [M_{\odot}]}{\tau [\rm \mu s]} = 
b_0 + b_1 \frac{M}{R_s} + b_2 \left( \frac{M}{R_s} \right) ^2 \ ,
\label{fits_R2}
\end{eqnarray}
shown by the grey curve for GR, the blue curve for $a=10$,
and the red curve for $a=10^5$.
Figure \ref{R2_scaled} shows, that these fits present 
a dependence with the coupling $a$.
The respective values of the fit parameters are given in 
\cite{Blazquez-Salcedo:2018qyy}.
The deviation of each data set from the corresponding universal relation 
given by the respective fit is typically of the order of 10\%, 
as seen in the lower panels of the figure. 

The phenomenological universal relations (\ref{fits_R2a})-(\ref{fits_R2}) 
show that the deviation from GR depends on the value of the compactness. 
This has some potential utility in order to derive constraints on the parameter 
$a$ of the theory and/or properties of the stars, 
once observations of neutron star ringdowns can be done with high enough accuracy. 
In an ideal scenario, the $a$ parameter could be constrained 
by future gravitational wave observations. 
In order to do so, in addition to the frequency and the damping time, 
the mass and radius of the star need to be known with high enough accuracy. 
With the knowledge of these quantities, it would then be possible 
to test the previous asteroseismology relations. 
From our current results we estimate that the frequency 
or the damping time need to be known with at least 10\% accuracy in 
order to be able to test deviations from GR.

But even if the precision is not sufficient to test deviations from GR, 
the universal relations could be useful in order to constrain 
the stellar parameters of a star. 
For instance, with the ratio between $\omega_R$ and $\omega_I$, 
one could impose constraints on the compactness of the star, 
which is related to the EOS.

Let us emphasize that so far we have only considered the case of axial modes. 
However, a full study of the asteroseismology relations has to include 
the polar part of the spectrum as well
\cite{Ferrari:2007dd,Andersson:1997rn,Kokkotas:1999mn,Benhar:2004xg}. 
This part of the spectrum will presumably include new modes 
related to the presence of a scalar field 
(in a way, similar to the results discussed for black holes 
in the previous section). 
{A combination of the universal relations of axial and polar modes, 
including scalar modes, could then be used to constrain 
both the range of the parameter of the theory and the EOS.}

{The polar quasinormal modes and more specifically the fundamental
$f$-modes of neutron stars in $R^2$-gravity
were studied until now only in the Cowling approximation \cite{Staykov:2015cfa}. 
Even though this approximation leads to large deviations in the QNM frequencies, 
it simplifies the problem a lot and can give
us a good intuition about the qualitative and even quantitative
differences between GR and the corresponding alternative theory of gravity. 
The results in \cite{Staykov:2015cfa} show that the differences in the
mode oscillation frequencies coming from the 
$R^2$ modification of gravity are non-negligible and larger than the equation
of state uncertainties, but still too small to be observed even by the current gravitational wave detectors. 
Of course, in order to obtain the full picture of the problem one should drop 
the Cowling approximation and consider the metric and the scalar field 
perturbations as well. This is work in progress.}

\subsection{Scalar-tensor theories} \label{sec_STT}

Let us now present results on the quasinormal modes
of neutron stars obtained in a different class of theories,
leading to scalarized neutron stars \cite{Damour:1993hw}.
While the model we consider is similar to (\ref{eq:action_R2}), with
\begin{equation}
S= \frac{1}{16\pi G}\int d^4x \sqrt{-g} \Big[ 
R - 2\nabla_\mu \phi \nabla^\mu \phi
+ L_{matter}(A^2(\phi) g_{\mu\nu},\xi) 
\Big] ,
\label{action_Einstein}
 \end{equation}
it does not have a potential for the scalar field
and the coupling function $A(\phi)$ is chosen
to possess a quadratic next-to-leading order term,
\begin{equation}
{A}(\phi)=e^{\frac{1}{2}\beta\phi^2} \ . 
\label{A}
\end{equation}

\begin{figure}[h]
	\centering
	\includegraphics[width=0.38\textwidth,angle=-90]{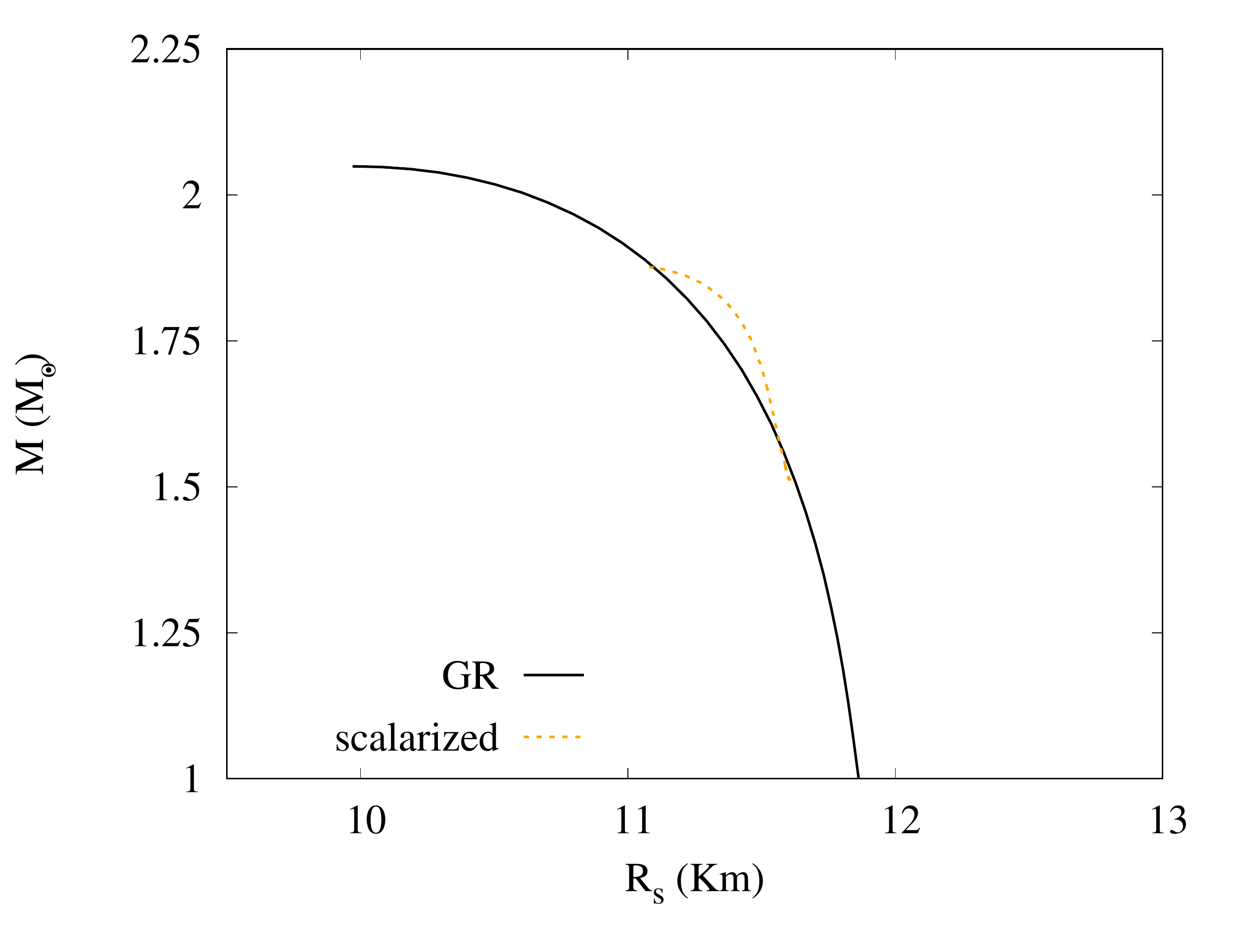}

	\caption{Mass-radius relation for neutron stars in STT gravity
for EOS Sly for $A=e^{\frac{1}{2}\beta\phi^2}$
with $\beta=-4.5$;
black: GR, orange: scalarized.}
	\label{STT_MR}
\end{figure}

In this case the GR neutron stars remain solutions of the theory. 
However, similarly to the case of the scalarized black holes 
discussed in subsection \ref{sec_sEGB}, 
neutron stars with a non-trivial scalar field bifurcate at certain points 
from the GR solutions. 
This is known as spontaneous scalarization of neutron stars 
\cite{Brans:1961sx,Damour:1992we,Damour:1993hw,Damour:1996ke,Fujii:2003pa}. 
Various properties of spontaneously scalarized neutron stars 
have been studied in the literature 
\cite{Damour:1993hw,Damour:1996ke,Harada:1998ge,Salgado:1998sg,Sotani:2012eb,Pani:2014jra,Sotani:2017pfj,Doneva:2013qva,Doneva:2014uma,Doneva:2014faa,Staykov:2016mbt,Yazadjiev:2016pcb,Doneva:2016xmf}.
In Fig.~\ref{STT_MR} we present a typical mass-radius diagram 
for neutron stars in the above theory for $\beta=-4.5$,
and employing the SLy EOS.

\begin{figure}[h]
\begin{center}
\vspace*{-0.5cm}
{\includegraphics[width=0.49\textwidth, angle =-90]{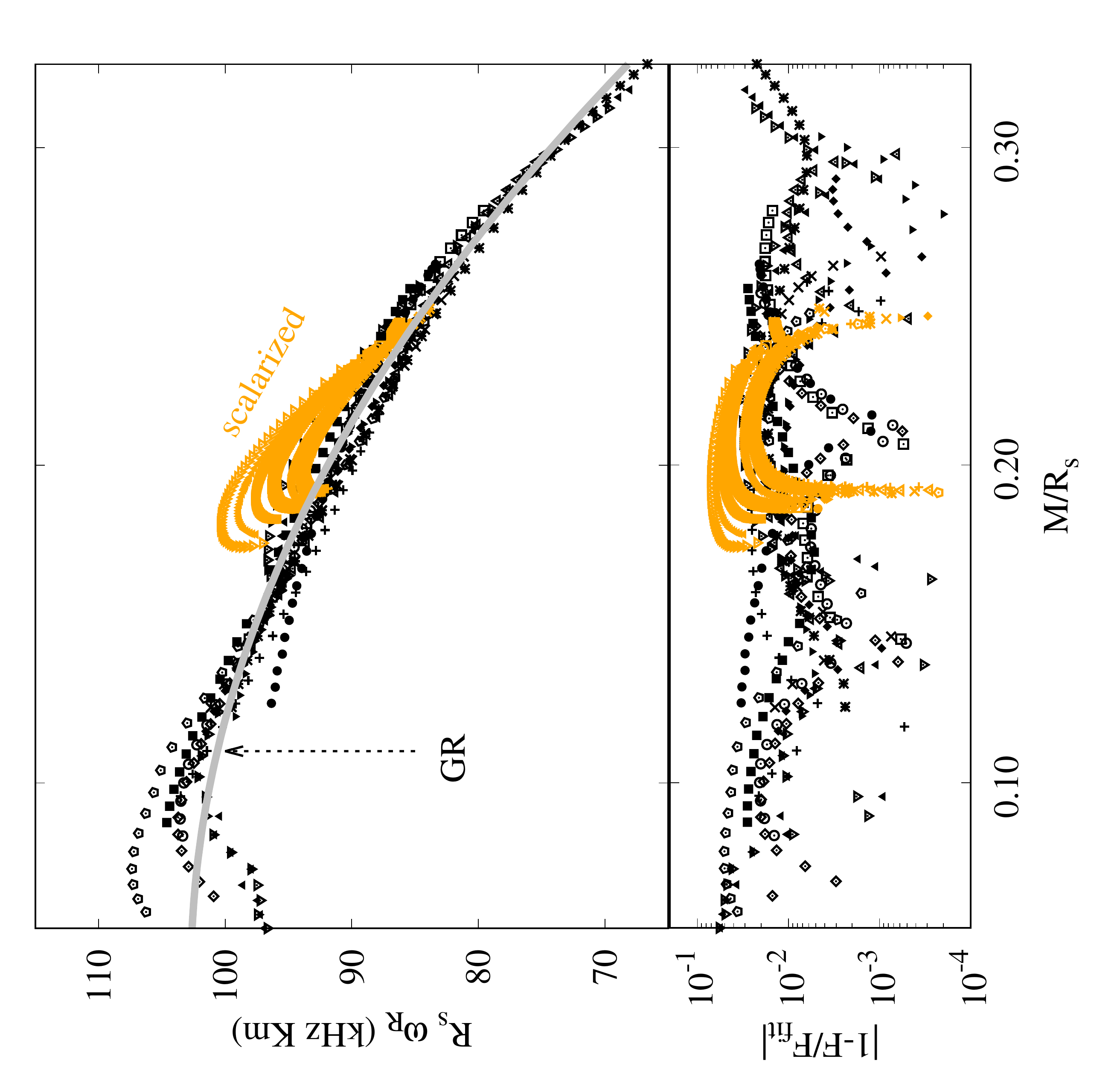}}
{\includegraphics[width=0.49\textwidth, angle =-90]{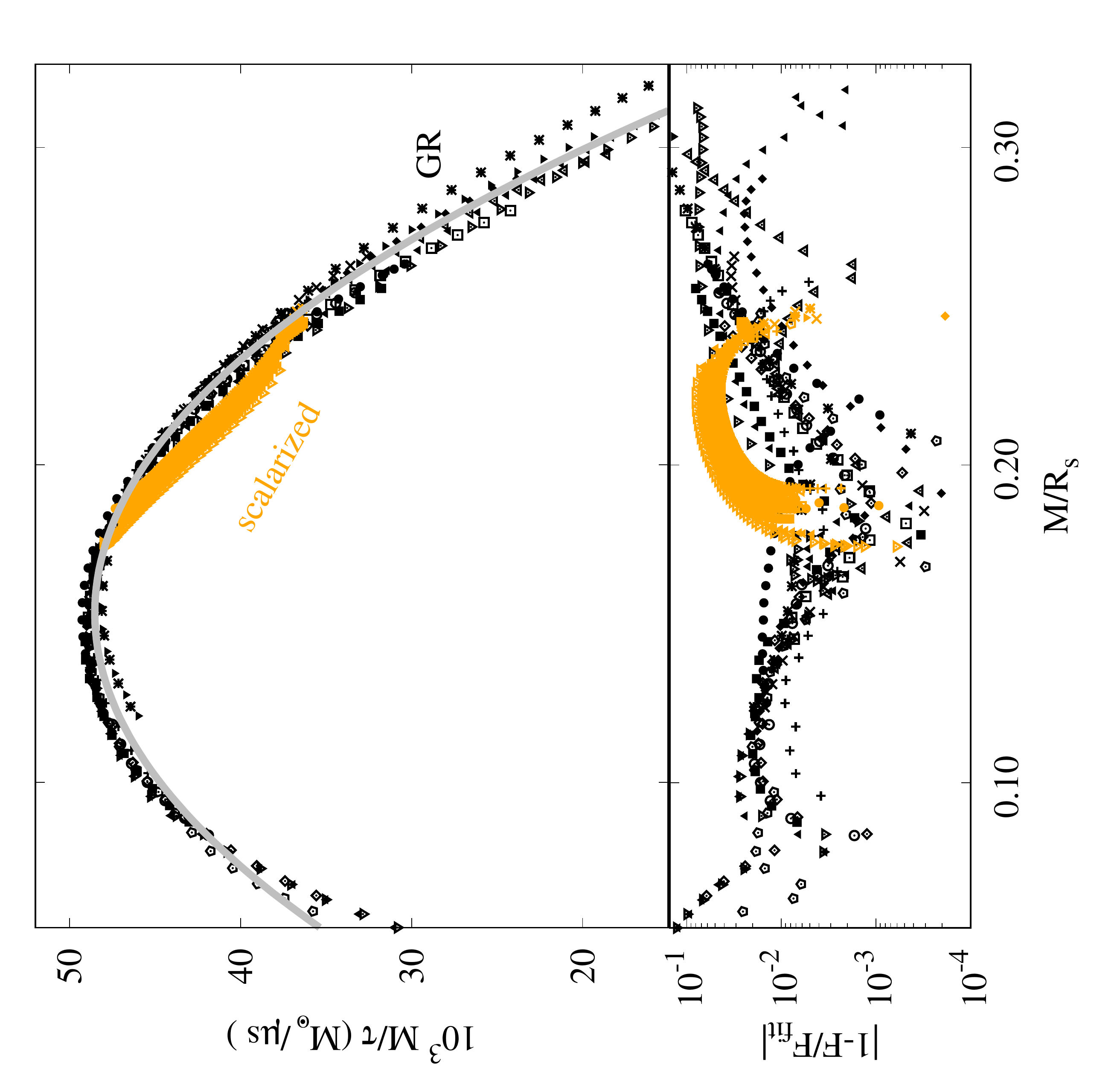}}
\end{center}
\caption{(left)
The frequency $\omega_{R}$ scaled by the radius $R_s$ of the star
(in $\rm kHz \cdot km$) vs the compactness $M/R_s$
for neutron stars in STT theory.
(right)
Inverse of the damping time $\tau$ scaled by the mass 
(in $M_\odot/\mu$s) vs $M/R_s$.
Fourteen different EOSs are shown (see \cite{AltahaMotahar:2018djk})
for GR configurations (black) and the scalarized solutions
with $A=e^{\frac{1}{2}\beta\phi^2}$
and $\beta=-4.5$ (orange).
The solid lines (grey) correspond to a quadratic fit.
In the bottom panels the difference between the fits and the data
is shown, with $F$ denoting the corresponding scaled quantity
in the panel above.
}
\label{axial_STT}
\end{figure}

The axial quasinormal modes of the neutron star models
obtained with the above coupling function $A(\phi)$ for several
values of $\beta$ 
{have been studied for the first time in 
\cite{Sotani:2005qx}. These results were later 
extended in \cite{AltahaMotahar:2018djk} for an alternative 
`quadratic' coupling function. 
In addition to all the EOSs employed in the previous subsection
for $R^2$ gravity, several more EOSs have been employed,}
corresponding to three new equations proposed recently
by Paschalidis et al.~\cite{Paschalidis:2017qmb} 
for hybrid nuclear-quark stars, 
and a simple polytropic EOS \cite{AltahaMotahar:2018djk}.

The results for the axial $l=2$ quasinormal modes 
are shown in Fig.~\ref{axial_STT}, 
which includes all the data for the various EOSs, 
presenting the GR results (black), 
and the STT results (orange) for $\beta=-4.5$. 
This value of $\beta$ is close to the maximum value 
allowed by the observations of the binary pulsar PSR J1738+0333 
\cite{Freire:2012mg} for massless scalar fields \cite{Archibald:2018559}.

Interestingly, the axial quasinormal modes are not very sensitive 
to the presence of scalarization in the neutron stars. 
Note, that in Fig.~\ref{axial_STT} both the scaled frequency 
and the scaled damping time do not deviate strongly from the GR results. 
In other words, the variation of the modes (scaled or unscaled) 
resulting from changing the EOS
is of the same order of magnitude as the variation 
introduced by the presence of a scalar field. 
The particular values and forms of the universal relations 
can be found in \cite{AltahaMotahar:2018djk} 
(here universal relations between the modes and the moment of inertia of the neutron stars are also studied with similar results).

In this case the universal relations cannot realistically allow us 
to distinguish between GR and scalarized neutron stars. 
However, since the universal relations are rather independent of the theory, 
they can be used to obtain information on the properties of neutron stars.
For example, from the ratio $\omega_R/\omega_I$
the compactness of the star could be derived \cite{Tsui:2004qd}. 
This could potentially be used together with additional measurements 
of the mass or radius of a star to constrain the EOS.

{The polar $f$-modes of neutron stars in scalar-tensor theories of gravity were
examined in \cite{Sotani:2004rq} in the nonrotating case. As a matter of fact this was 
the first study of QNMs of neutron star in alternative theories of gravity
in general. The calculations were performed in Cowling approximation, similar 
to the $R^2$-gravity case discussed above. The results show that the frequencies of scalarized
neutron stars can deviate a lot from pure GR for small enough values of $\beta$, 
but if one imposes the observational constraints, i.e. $\beta>-4.5$, the differences
are very small, within the equation of state uncertainties, similar to the case of axial 
modes.}

{Another interesting case of neutron star oscillations in scalar-tensor theories of 
gravity was considered in \cite{Silva:2014ora} where the torsional oscillations
were examined. The studies showed that the effect of scalarization on the
mode frequencies is again smaller than the uncertainties from the microphysics. Thus, 
the observation of quasi-periodic oscillations following giant flares can be used 
in order to constrain the neutron star crust models, independently of whether 
spontaneous scalarization occurs or not.}

{From the above overview of the literature one can conclude that for the values of 
$\beta$ that are in agreement with the binary pulsar observations, the non-rotating neutron star 
QNM  frequencies  and damping times cannot be used to set further constraints on 
the scalar-tensor theories of gravity since the deviations from pure GR are 
completely within the EOS uncertainties. This is not the case, though, if we
consider rapid rotation close to the Kepler (mass shedding) limit. Then the
scalarization can lead to non-negligible effects in the neutron star properties even for 
$\beta>-4.5$ \cite{Doneva:2013qva}. With this motivation in mind the $f$-mode oscillation 
frequencies and damping times of rapidly rotating neutron stars were studied in the Cowling 
approximation \cite{Yazadjiev:2017vpg} with a special emphasis on the rotational driven 
Chandrasekhar-Friedman-Schutz (CFS) instability \cite{Chandrasekhar:1992pr,Friedman:1978hf}, 
that is a secular instability which develops due to the emission of gravitational waves. The results
showed that the presence of nontrivial scalar field indeed influences significantly
the $f$-mode frequencies and damping times even for $\beta>-4.5$. On one hand the 
scalarization facilitates the development of the CFS instability by reducing the threshold value 
of the normalized angular momentum where this instability starts to operate but on the other hand 
the growth time of the instability increases compared to pure general relativity which hinders
the gravitational wave emission. }

\subsection{Non-minimal derivative coupling in Horndeski gravity} \label{sec_Hg}

So far we have considered neutron stars in a particular type of STT. 
However, one can also go to a set of more general STTs,
containing only second order derivatives in the field equations, 
possessing no ghosts and satisfying global hyperbolicity, 
as formulated by Horndeski \cite{Horndeski:1974wa},
and related to covariant Galileon gravity in four dimensions 
\cite{Nicolis:2008in,Kobayashi:2011nu}. 
Neutron stars in Horndeski and beyond Horndeski gravity 
have been studied in the literature  
in \cite{Maselli:2016gxk,Babichev:2016jom,Babichev:2016rlq}.

One particular set of such theories is known as the 
non-minimal derivative coupling sector of Horndeski theory. 
The post-Newtonian analysis reveals 
that this set of Horndeski theories is very constrained by solar system tests 
\cite{Bruneton:2012zk}. 
In the special case addressed here, the corresponding action then reduces to
\cite{Blazquez-Salcedo:2018tyn}
\begin{eqnarray}
\label{action_nmdch} 
S=\frac{1}{16\pi G}\int  d^4x  \sqrt{-g} \Big[ R 
+ 2 G^{\mu\nu} \nabla_{\mu}\phi \nabla_{\nu}\phi + L_{matter} \Big]
,
\end{eqnarray}
where $G^{\mu\nu}$ is the Einstein tensor.

\begin{figure}[h]
	\centering
	\includegraphics[width=0.38\textwidth,angle=-90]{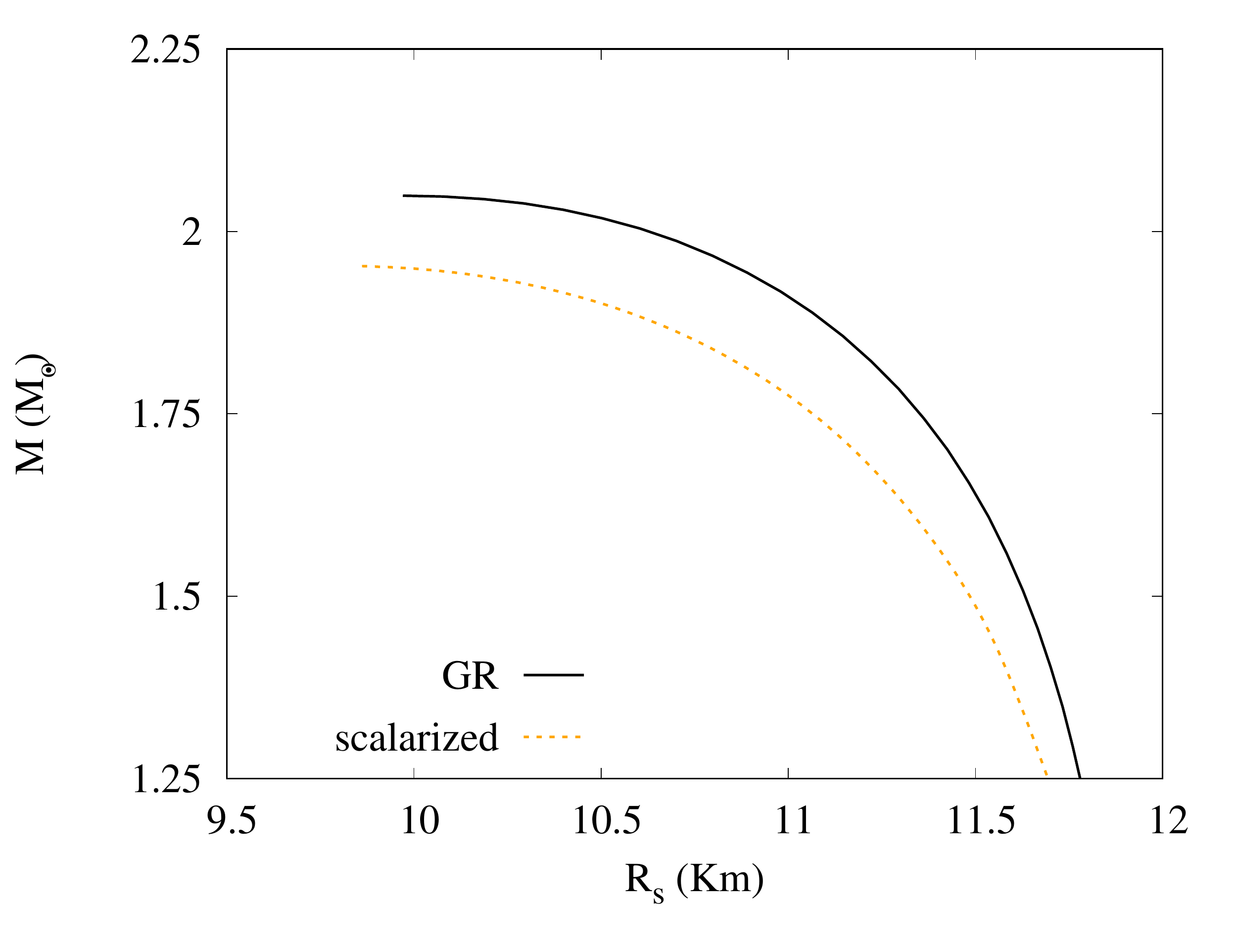}
	
	\caption{Mass-radius relation for static neutron stars in Horndeski
theory with non-minimal derivative coupling for EOS SLy; black solid: GR,
orange dotted: scalarized.}
	\label{nmdch_MR}
\end{figure}

Interestingly, the theory possesses asymptotically flat neutron stars 
that could be astrophysically relevant \cite{Cisterna:2015yla,Cisterna:2016vdx}.
Outside the neutron star the scalar field vanishes,
and the metric coincides with a GR vacuum solution. 
But the interior of the neutron star contains a non-trivial scalar field 
that couples to gravity and matter. 
This modifies the matter distribution 
and the global properties of the configuration. 
In Fig.~\ref{nmdch_MR} we show a typical mass-radius relation 
for static and spherically symmetric configurations
in the above Horndeski model for the EOS Sly,
and compare it to the corresponding GR relation
\cite{Blazquez-Salcedo:2018tyn}.
Note, that the deviation in terms of mass and radius 
from GR is not very large.

However, the quasinormal mode analysis of the axial channel reveals 
that the spectrum of the star is much more strongly influenced 
by the presence of the scalar field \cite{Blazquez-Salcedo:2018tyn}.
For instance, the damping times of the scalarized stars with the largest masses 
are considerably shorter than the corresponding damping times 
of GR configurations.

\begin{figure}[h]
\begin{center}
\vspace*{-0.5cm}
{\includegraphics[width=0.49\textwidth, angle =-90]{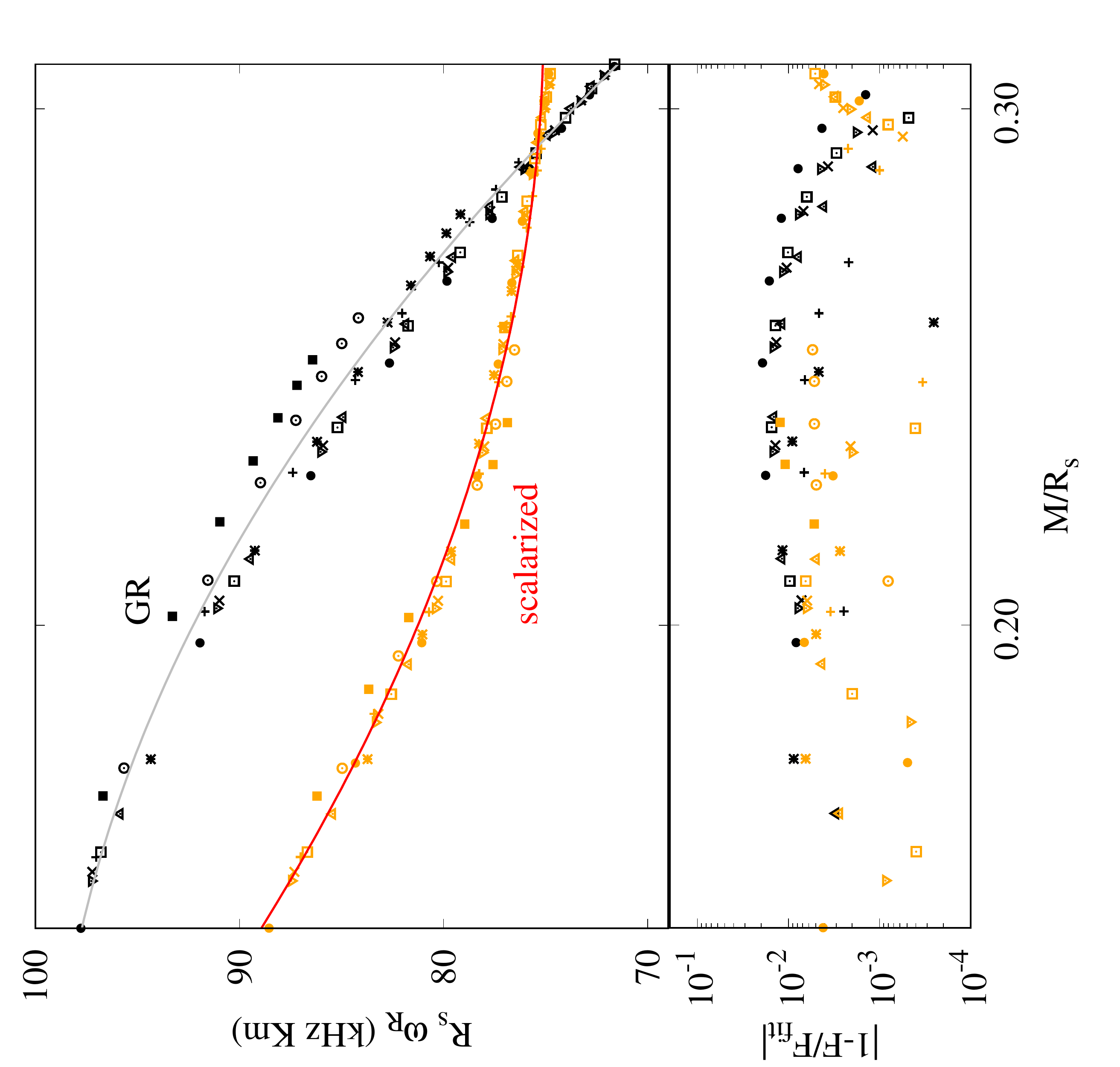}}
{\includegraphics[width=0.49\textwidth, angle =-90]{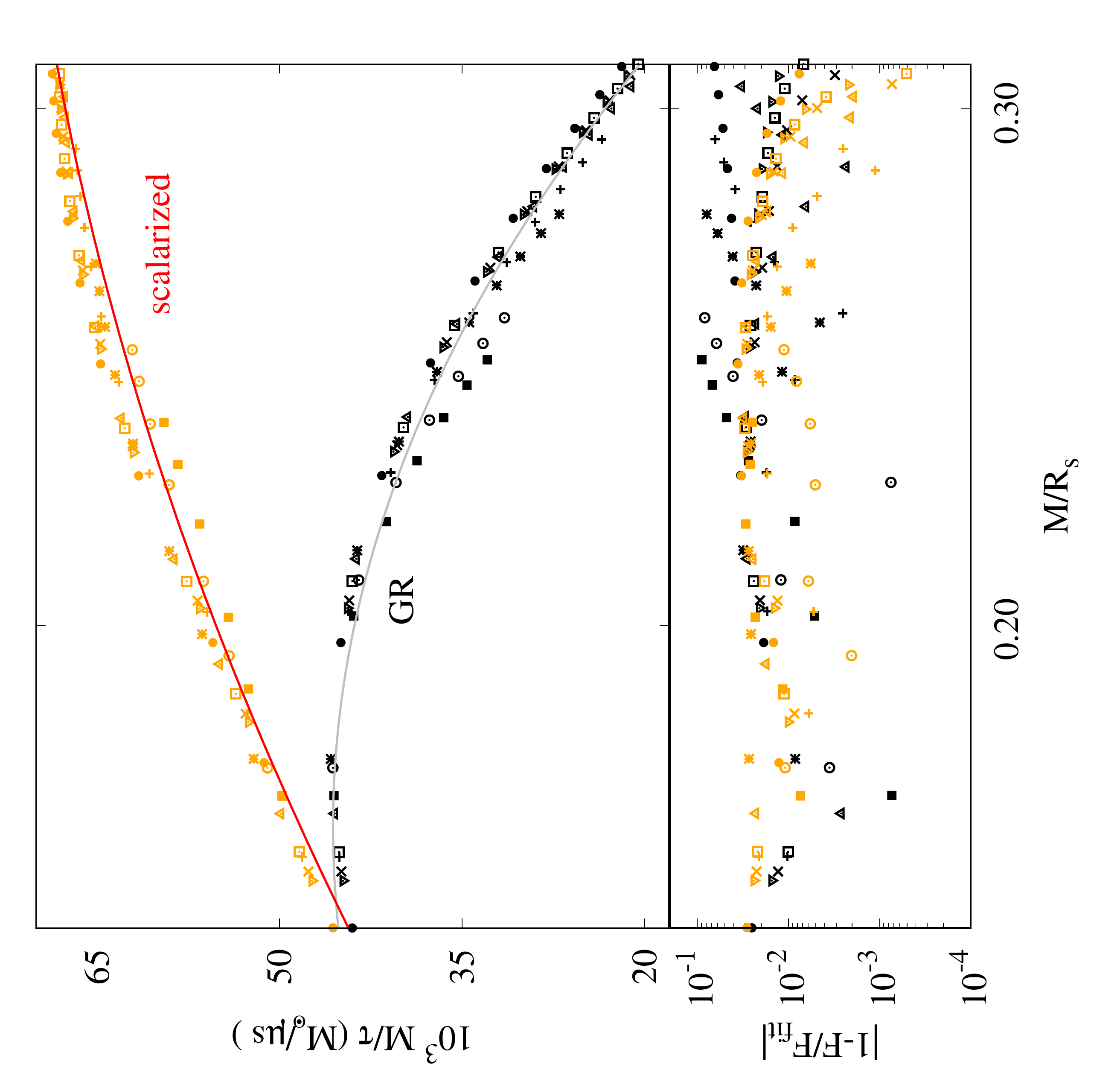}}
\end{center}
\caption{(left)
The frequency $\omega_{R}$ scaled by the radius $R_s$ of the star
(in $\rm kHz \cdot km$) vs the compactness $M/R_s$
for neutron stars in the minimally coupled sector of Horndeski theory.
(right)
Inverse of the damping time $\tau$ scaled by the mass
(in $M_\odot/\mu$s) vs $M/R_s$.
Thirteen different EOSs are shown (see \cite{Blazquez-Salcedo:2018tyn})
for GR configurations (black) and the scalarized solutions
(orange).
The solid lines correspond to a quadratic fit for GR (grey)
and the scalarized solutions (red).
In the bottom panels the difference between the fits and the data
is shown, with $F$ denoting the corresponding scaled quantity
in the panel above.
}
\label{nmdch_axial}
\end{figure}

In Fig.~\ref{nmdch_axial} we show the universal relations for the axial
quasinormal modes for this particular Horndeski theory.
The left panel demonstrates that the scaled frequency 
of the scalarized stars tends to deviate considerably
from its GR counterpart for low values of the compactness. 
The right panel demonstrates that the scaled damping time
of the scalarized stars deviates strongly from 
its GR counterpart for high values of the compactness.
Thus the frequency and damping time possess 
very different universal relations for scalarized neutron stars
in this Horndeski theory as compared to GR.

Hence we conclude that in this particular model 
the presence of a scalar field has a more pronounced effect 
than in the above discussed STT model, 
even though the scalar field is now only localized 
inside the neutron star, 
while vanishing completely outside the star 
and modifying the mass-radius relation only slightly.

\subsection{Dilatonic-Einstein-Gau\ss -Bonnet theory} \label{sec_dEGB_ns}

As a final example, let us come back to the 
dEGB theory with action (\ref{sEGB_action}) and coupling (\ref{couling_dEGB}). 
This theory also supports neutron stars with a dilatonic field. 
Such neutron star models have been studied in detail in 
\cite{Pani:2011xm,Kleihaus:2014lba,Kleihaus:2016dui}. 

\begin{figure}[h]
	\centering
	\includegraphics[width=0.38\textwidth,angle=-90]{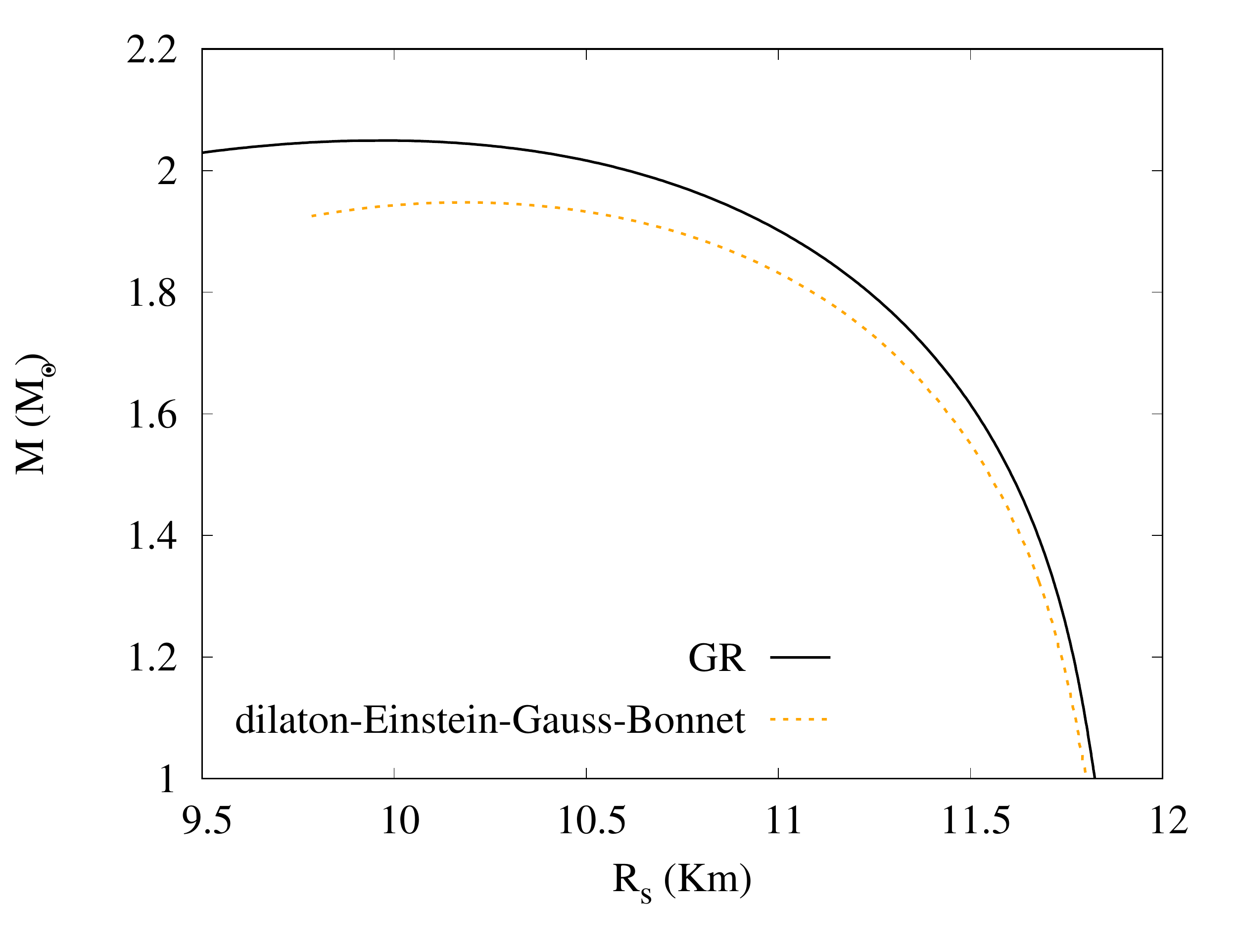}

\caption{Mass-radius relation for neutron stars in dEGB theory with 
coupling constants $\alpha = 0.268\times 10^{5}$ cm and $\beta=1$
for EOS SLy; black solid: GR,
orange dotted: dEGB.}
	\label{dEGB_MR}
\end{figure}

In Fig.~\ref{dEGB_MR} we show the typical mass-radius relation 
for neutron stars in this theory, employing the coupling constants
$\alpha = 0.268\times 10^{5}$ cm and $\beta=1$, and comparing to GR.
As shown before, the presence of the dGB term
leads to a decrease of the maximum mass of the neutron star models.

The axial quasinormal modes of these configurations have been studied in 
\cite{Blazquez-Salcedo:2015ets}, employing 
the coupling constants $\alpha = 0.268\times 10^{5}$ cm and $\beta=1$,
i.e., values within the current constraints for the theory. 
The analysis has shown that the presence of the dGB term
leads to an increase of the frequency and of the damping time 
with respect to configurations of similar mass in GR.

\begin{figure}[h]
\begin{center}
\vspace*{-0.5cm}
{\includegraphics[width=0.38\textwidth, angle =-90]{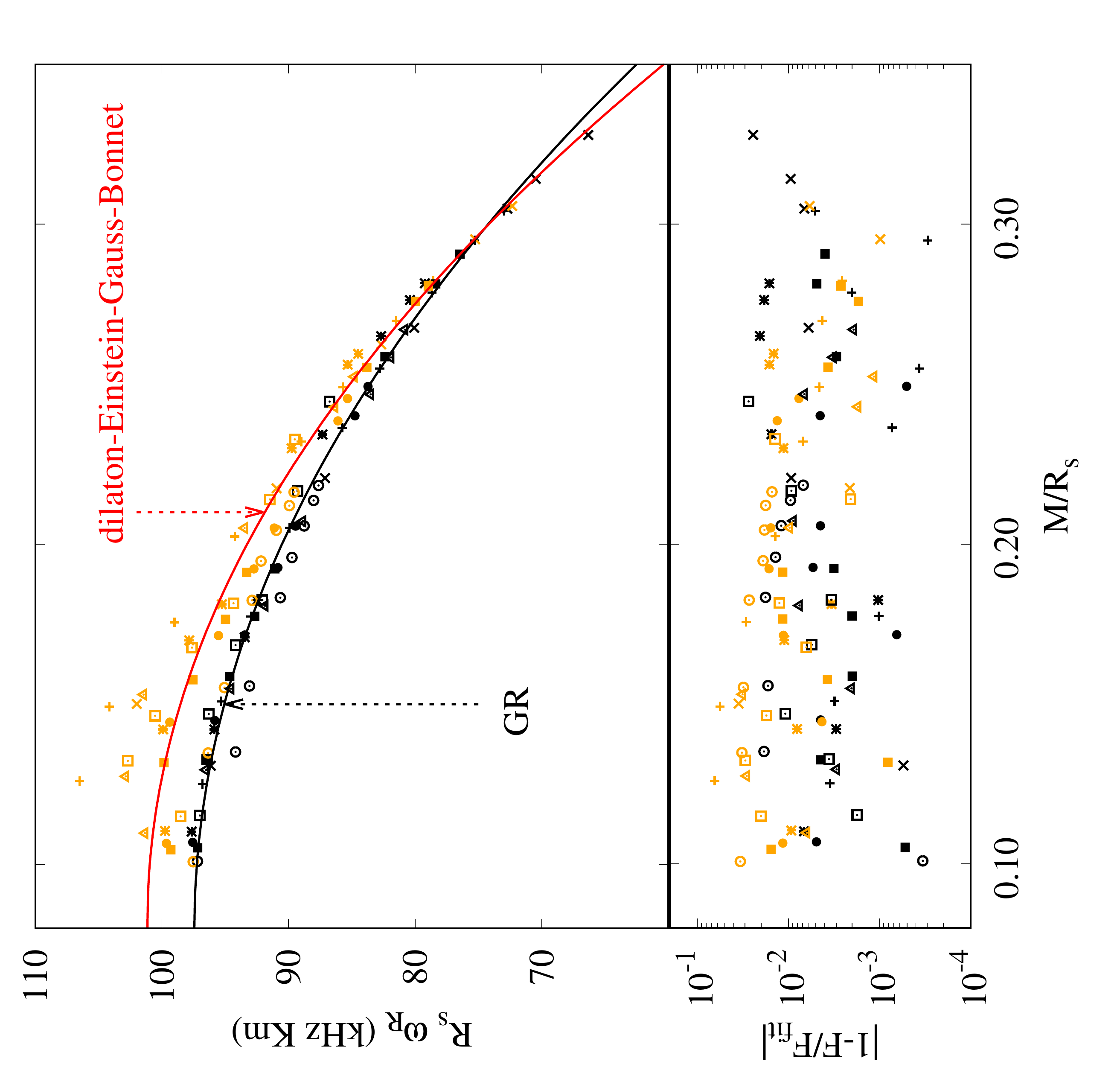}}
{\includegraphics[width=0.38\textwidth, angle =-90]{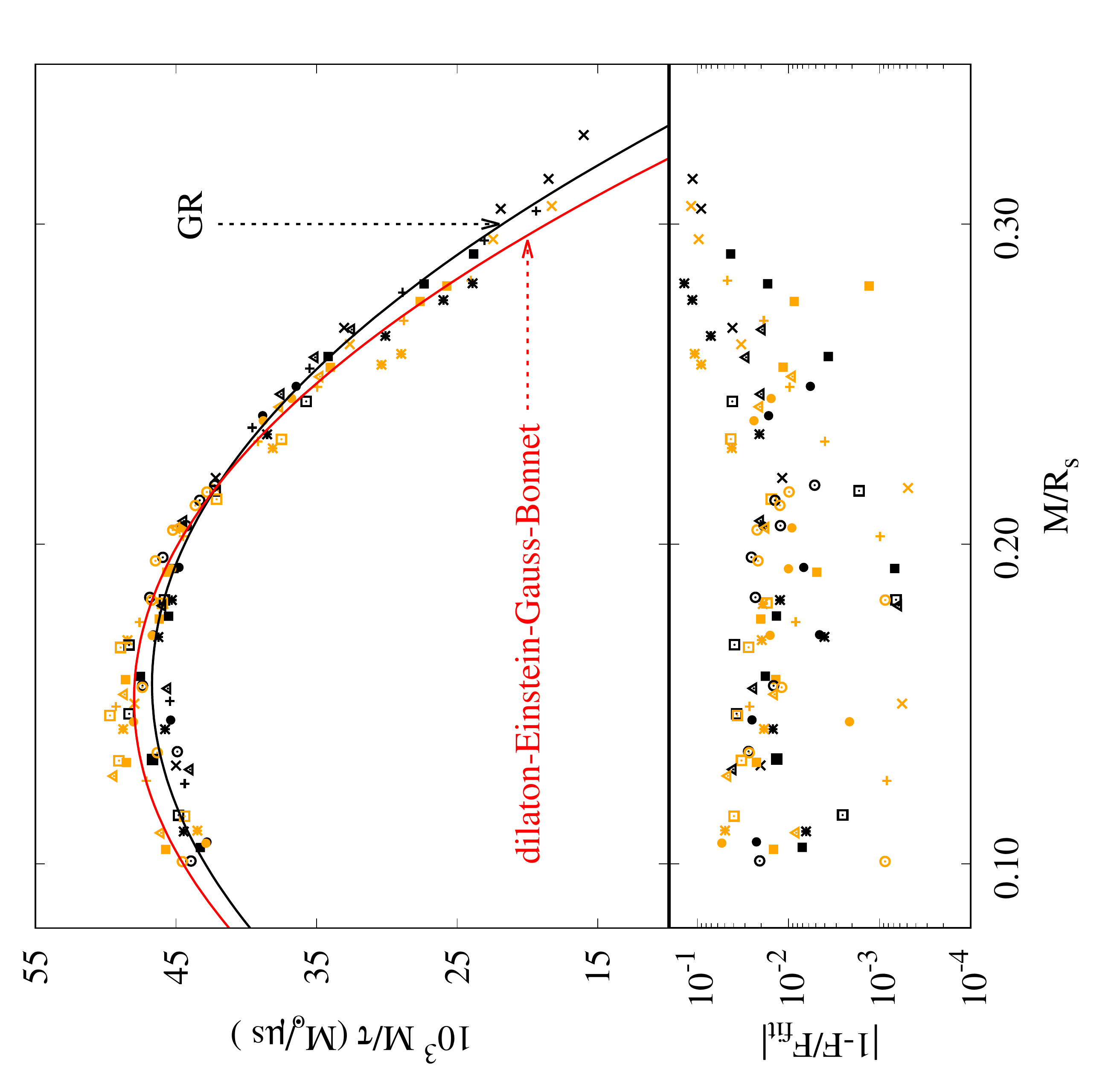}}
\end{center}
\caption{(left)
The frequency $\omega_{R}$ scaled by the radius $R_s$ of the star
(in $\rm kHz \cdot km$) vs the compactness $M/R_s$
for neutron stars in dEGB theory
with coupling constants
$\alpha = 0.268\times 10^{5}$ cm and $\beta=1$.
(right)
Inverse of the damping time $\tau$ scaled by the mass
(in $M_\odot/\mu$s) vs $M/R_s$.
Eight different EOSs are shown (see \cite{Blazquez-Salcedo:2015ets})
for GR configurations (black) and the dEGB solutions (orange).
The solid lines correspond to a quadratic fit for GR (grey)
and the scalarized solutions (red).
In the bottom panels the difference between the fits and the data
is shown, with $F$ denoting the corresponding scaled quantity
in the panel above.
}

\label{axial_dEGB}
\end{figure}

In Fig.~\ref{axial_dEGB} we show the results for the scaled $l=2$ 
fundamental axial modes. 
Here eight different EOSs are shown (see \cite{Blazquez-Salcedo:2015ets}),
both for dEGB and GR neutron stars.
This result is somewhat similar to the one in subsection \ref{sec_STT}: 
the deviations in the universal relations for the different theories 
are of the same order of magnitude as the deviations 
due to a change in the EOS.

\section{Conclusions and outlook}

Compact astrophysical objects like black holes and neutron stars
represent valuable testing grounds for alternative theories of gravity,
since highly accurate observations of the physics 
in strong gravitational fields
effectively augment solar system observations.
Of particular importance here are the recent and future
observations of gravitational waves together with the associated
multi-messenger astronomy.

The detection of gravitational waves emitted in the ringdown of a 
compact object, that has been created via a merger,
offers a particularly promising tool to improve our understanding
of gravity. 
Whereas GR predicts a relatively simple set of quasinormal modes
of black holes, allowing only for quadrupole radiation,
and showing in the case of static black holes even the phenomenon
of isospectrality, and thus the same mode spectra for
axial and polar modes, this may be very different for
alternative theories of gravity.

Here we have addressed the quasinormal mode spectra for
several alternative theories of gravity, which all include
a scalar field, either from the beginning or after a reformulation.
The presence of a scalar field, however, implies the
emission of monopolar and dipolar radiation in addition to
quadrupole radiation in the ringdown of black holes. Moreover, 
isospectrality is broken. We have demonstrated this
by evaluating the normal mode spectra of dEGB black holes.

The quasinormal mode spectra of
neutron stars may appear in this connection, at first sight,
as a less efficient means to draw conclusions on gravity theories,
since the EOS dependence of the properties of neutron stars
also concerns their quasinormal modes.
However, the existence of universal relations of neutron star
frequencies and damping times makes them still potentially
rather useful to learn about various gravity theories.

Our analysis of quasinormal modes of neutron stars in
alternative theories of gravity has so far focused on the
axial modes, which are much easier to obtain than the polar
modes, since the perturbations of the scalar field 
and the nuclear matter are decoupled. Thus only the metric
perturbations alone must be analyzed.
But this analysis has already brought forward some interesting 
observations for some of the theories studied here.
For instance, the universal relations show a quite noticable
dependence on the parameter $a$ in $R^2$ theory,
and a very notable differing dependence in Horndeski theory.

Certainly, the next step is to thoroughly 
analyze the quasinormal modes for the case of the polar modes.
Here, as in the case of the black holes, also for the neutron stars
new channels due to the scalar field will be present. 
However, since the polar modes include the perturbations of
the stellar matter, a large variety of effects and deviations
from the known modes of GR may arise in addition.
Together these studies may then shed light both on the
gravity theories and on the EOS.

Further into the future, we envisage to extend these investigations
of quasinormal modes
to the case of rotating black holes and neutron stars.
The presence of rotation will
make these studies technically considerably more
challenging, but on the other hand it will allow 
a much wider application concerning astrophysical 
and gravitational wave observations, 
which typically include rotating compact objects.

\section*{Acknowledgements}
JLBS would like to acknowledge support from the DFG project BL 1553. 
JLBS, ZAM and JK would like to acknowledge support by the 
DFG Research Training Group 1620 {\sl Models of Gravity} 
and the COST Action CA16104.
FSK acknowledges support by the Croatian Science Foundation under the Project IP-2014-09-3258 and
the H2020 Twinning Project No. 692194 “RBI-T-WINNING”.

\bibliographystyle{unsrt}

\end{document}